\documentclass[12pt,preprint]{aastex}

\newcommand\msun{\rm M_{\odot}}

\newcommand\rsun{\rm R_{\odot}}
\newcommand\msunyr{\rm M_{\odot}\,yr^{-1}}
\newcommand\be{\begin{equation}}
\newcommand\en{\end{equation}}

\newcommand\mdot{\dot{M}}

\newcommand\curf{{\cal F}}

\newcommand\ha{H$\alpha \;$}

\begin{document}
\title{Accretion in Young Stellar/Substellar Objects}
\author{
James Muzerolle\altaffilmark{1,2},
Lynne Hillenbrand\altaffilmark{2,3},
Nuria Calvet\altaffilmark{4},
C\'esar Brice\~no\altaffilmark{2,5},
Lee Hartmann\altaffilmark{4}
\altaffiltext{1}{Steward Observatory, 933 N. Cherry Ave., University of Arizona,
Tucson, AZ 85721}
\altaffiltext{2}{Visiting Astronomer, W. M. Keck Observatory, which is
operated as a scientific partnership among the California Institute of
Technology, the University of California, and the National Aeronautics and
Space Administration.  The Observatory was made possible by the generous
financial support of the W. M. Keck Foundation.}
\altaffiltext{3}{Dept. of Astronomy, MS105-24, California Institute of Technology,
Pasadena, CA 91125}
\altaffiltext{4}{Harvard-Smithsonian Center for Astrophysics, 60
Garden St., Cambridge, MA 02138}
\altaffiltext{5}{Centro de Investigaciones de Astronomia (CIDA),
Apartado Postal 264, M\'erida 5010-A, Venezuela}}

\begin{abstract}
We present a study of accretion in a sample of 45 young,
low mass objects in a variety of star forming regions and young associations,
about half of which are likely substellar.  Based primarily on the presence
of broad, asymmetric \ha emission, we have identified 13 objects
($\sim$30\% of our sample) which are strong candidates for ongoing accretion.
At least 3 of these are substellar.  We do not
detect significant continuum veiling in most of the accretors with late
spectral types (M5-M7).  Accretion shock models show that lack of measurable
veiling allows us to place an upper limit to the mass accretion rates of
$\lesssim 10^{-10} \; \msunyr$.  Using magnetospheric accretion models
with appropriate (sub)stellar parameters, we can successfully explain
the accretor \ha emission line profiles, and derive quantitative estimates
of accretion rates in the range $10^{-12} < \mdot < 10^{-9} \; \msunyr$.
There is a clear trend of decreasing accretion rate with stellar mass,
with mean accretion rates declining by 3-4 orders of magnitude over
$\sim 1 - 0.05 \; \msun$. 
\end{abstract}

\keywords{accretion disks, brown dwarfs, stars: emission-line, pre-main sequence, circumstellar matter,}

\section{Introduction}

The collapse ($\sim10^5$ yr; Adams, Lada, \& Shu, 1987) of a slowly rotating
molecular cloud core leads to formation of a protostar and surrounding
circumstellar accretion disk which can last up to
$\sim10^7$ yr in at least some cases (Muzerolle et al., 2000a;
Alencar \& Batalha, 2002).  Observable signatures of disks and disk accretion
include far- and mid-infrared emission from dust at a range of temperatures,
near-infrared emission from hot dust and gas in the inner disk, and a variety
of emission lines and UV/optical continuum excess due to accretion
of material from the inner disk directly
onto the star (see, e.g. Mannings, Boss, \& Russell 2000 for reviews).

Historically, accretion from the disk onto the star was thought to occur
through a hot equatorial ``boundary layer" (e.g. Lynden-Bell \& Pringle
1974; Bertout, Basri, \& Bouvier 1988; Basri \& Bertout 1989).
Over the past decade, however, a new model of {\it magnetospheric} accretion
has been widely adopted for classical T Tauri stars (hereafter, CTTSs).
This model was originally discussed in the context of neutron stars
by Ghosh \& Lamb (1979) and first applied to CTTSs by
Uchida \& Shibata (1985), Bertout et al. (1988), and Konigl (1991).
In this scenario, the inner regions of the circumstellar disk are
disrupted by the stellar magnetosphere, which channels viscously
accreting material out of the disk plane and
towards the star along magnetic field lines; an accretion shock is
formed as the free-falling material impacts the stellar
surface. This model can also generate magnetocentrifugal outflows,
which are launched from a region just exterior to the point of disk truncation 
(Shu et al, 1994), or from near the poles of the star (Hirose et al. 1997).
Several of the detailed observational characteristics of CTTSs, such as
blue continuum excesses including the Paschen and Balmer continuum
slopes (e.g. Calvet \& Gullbring, 1998),
relatively small near-infrared excesses (e.g. Meyer, Calvet 
\& Hillenbrand, 1997), permitted emission-line profiles (e.g. Muzerolle, 
Calvet, \& Hartmann, 2001a; hereafter, MCH), magnetic field strengths
(Johns-Krull et al. 1999; Johns-Krull \& Valenti 2000), and slow rotation
periods (e.g. Edwards et al. 1993), are successfully explained
by such a model.

The presence of inverse P Cygni line profiles in the spectra of
many CTTSs motivated Calvet \& Hartmann (1992) and then
Hartmann, Hewett, \& Calvet (1994) to first investigate
magnetospheric accretion flows as the source of the line emission.
Radiative transfer calculations of 
magnetospheric accretion predict a characteristic blue asymmetry
-- with or without a redshifted absorption feature --
in both hydrogen (Balmer, Paschen, and Brackett lines)
and metallic (notably NaD, OI, and the CaII triplet) line profiles.
These calculations successfully reproduce both the line fluxes
and the line profiles (Hartmann, Hewett, \& Calvet 1994; Muzerolle et al.
1998ab; MCH) observed in most CTTSs in the Taurus molecular cloud.

Results thus far, however, have been limited to T Tauri stars of restricted
range in mass and age ($M \sim 0.5 \; M_{\odot}$; $\tau<$1-3 Myr ).
Few high-resolution observations even exist of stars with masses 
$<0.2 \; M_{\odot}$.  In order to develop a broader understanding of
accretion disk activity, in this contribution we extend the analysis 
of accretion to a mass range more representative of the stellar mass function,
focusing on young brown dwarfs and stars in the mass range 0.05-0.2 M$_\odot$
(hereafter referred to as very low mass, or ``VLM", objects).
Our results for a single object among this sample, V410 Anon 13,
appeared in Muzerolle et al. (2000b), the first measurement of accretion in a
candidate sub-stellar mass young T Tauri object.  The derived
$\mdot$ of $\sim5\times 10^{-12} \; \msunyr$ remains the lowest measured
accretion rate, even when the larger sample presented here is considered.
We have also recently extended the analysis in age to study $\sim$10 Myr old 
stars via an investigation of disk accretion in the TW Hya association
(Muzerolle et al 2000a; Muzerolle et al. 2001b).

We describe our sample in section 2 and our observations, 
moderate- to high-resolution optical spectra, in section 3. 
Section 4 presents results from the spectra, including \ha line profiles,
measurements of continuum veiling, and model-derived mass accretion rates,
as well as radial and rotation velocities for part of the sample.  Finally,
we discuss the implications of our results for accretion in young VLM objects
in section 5.

\section{Sample}

The sample was selected from previously identified young
low mass objects in several different star forming regions and young
clusters, including Taurus, IC 348, $\sigma$ Ori, Upper Sco, and $\rho$
Oph (see references in Table~\ref{info}).
Our observations were designed in early 2000 to include every known
object with spectral type M5 or later and age $\sim$10 Myr or younger.
Since our observations were obtained, many more such 
objects have been discovered in star forming regions and young clusters
(e.g. Brice\~no et al. 2002) and thus our sample is not complete,
though it is likely to be representative of these very low masses. 
The final observed sample encompasses 45 young stars and brown dwarfs
with spectral types M4-M8.5.  At the 0.1-8 Myr ages of these regions,
the masses inferred from pre-main sequence evolutionary tracks
(e.g. D'Antona \& Mazzitelli, 1998; Chabrier et al. 2000) 
span the substellar boundary, ranging from 0.02-0.14 M$_\odot$.

\begin{figure}
\plotone{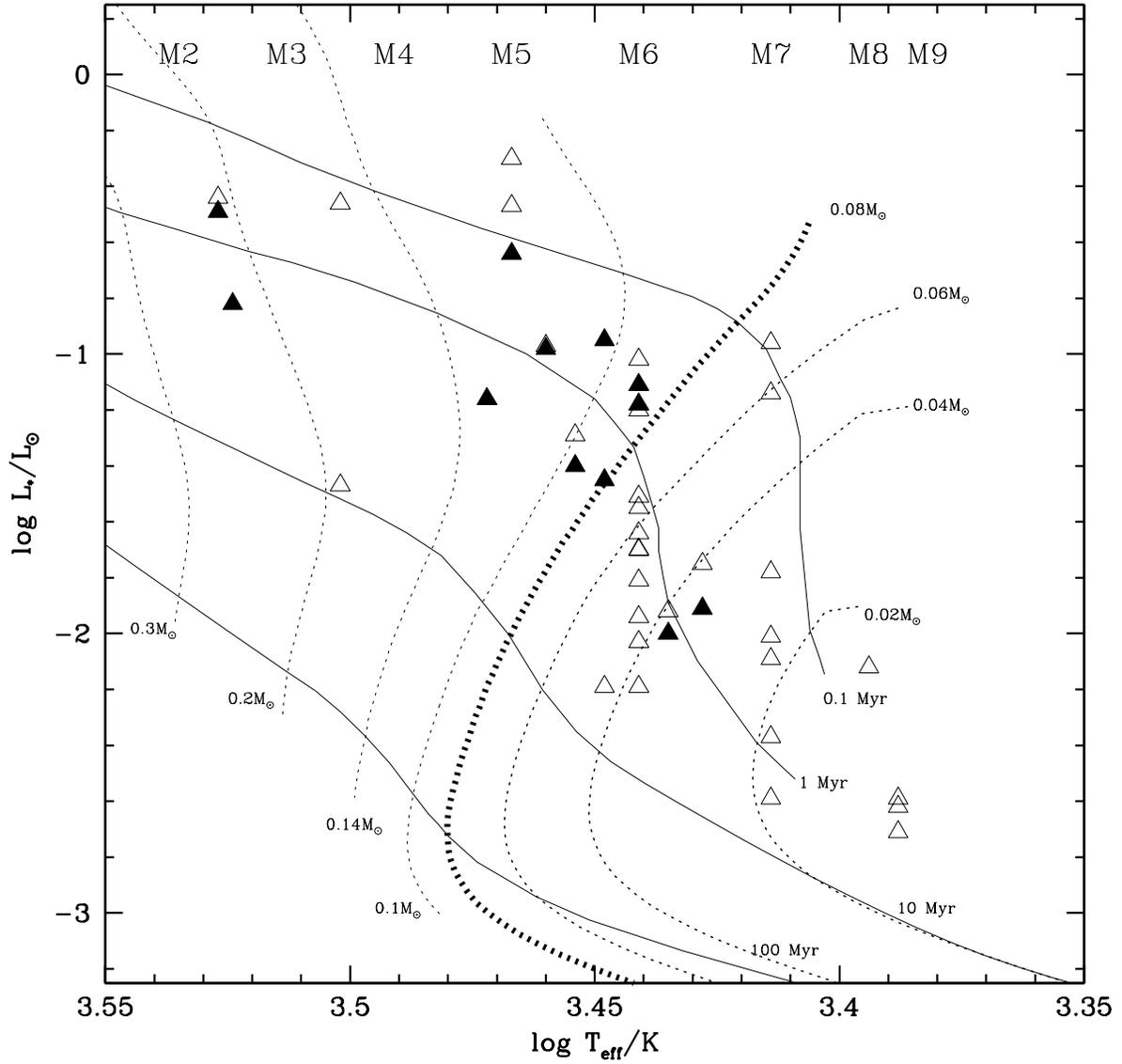}
\caption{HR diagram of entire sample, using pre-main sequence tracks
of D'Antona \& Mazzitelli (1998).  Solid triangles represent accretors,
open triangles represent non-accretors (see text).
\label{hrd}}
\end{figure}

Table~\ref{info}  contains basic information on our sample,
including spectral types
found in the literature from which, in combination with literature
photometry, the other parameters (logT, logL, extinction, H-K excess)
have been calculated.  Masses and ages were determined from the D'Antona
\& Mazzitelli (1997, including unpublished updates provided in 1998)
pre-main sequence evolution tracks, adopting the
transformations used in Hillenbrand \& Carpenter (2000).  
Our observations do not include enough 
dwarf and giant spectral templates to re-examine with any great
accuracy the previously published spectral types.
We note that one particular object, Lk\ha 358, has 
types from Cohen \& Kuhi (1979; M5.5) and Kenyon et al.
(1998; M5-M6) that are clearly {\it inconsistent} with our spectrum, 
which appears to be more like K8.  This star also displays spatially
extended H$\alpha$ and [SII] doublet emission, likely originating in a jet.
Our spectral type estimate of another object, Haro 6-28,
is also discrepant with the previously reported spectral type of M5
(Cohen \& Kuhi 1979); we estimate M2.5.  Both of these objects exhibit
accretion signatures, including continuum veiling.
Figure \ref{hrd} shows the HR diagram for our sample, and illustrates
the estimated mass and age ranges spanned in our study of accretion 
diagnostics.

\section{Observations}

We obtained spectra at Keck II with the ESI echellette spectrograph 
(Sheinis et al. 2002) on Jan. 25-28, 2001, and at Keck I with the HIRES echelle 
spectrograph (Vogt et al. 1994) on Jan. 10-11 2000 and June 30, 2001.  
Extraction of one dimensional spectra from the two-dimensional images
was performed using the MAKEE software written by T. Barlow.

ESI provides wavelength coverage from
3900 - 11000 \AA, but with only moderate resolution, $R \sim 8000$
(37.5 km s$^{-1}$), while HIRES has a smaller wavelength range, for our setup,
6320 - 8730 {\AA} with gaps between the orders and typical echelle resolution
of $R \sim 34,000$ (8.8 km s$^{-1}$).
All sets of data cover the important \ha emission line, our primary
diagnostic of accretion.  Despite lower resolution, the ESI data still 
resolve this line when accretion is present, since in that case the line widths
are typically several hundred km s$^{-1}$; however, chromospheric emission line
profiles and photospheric atomic absorption features generally are not resolved.
The trade-off for lower resolution is the benefit of 
increased wavelength coverage into the blue which yields a larger
baseline for measuring continuum veiling (although in most of our objects,
the S/N is too low to make measurements below $\sim 5000$ {\AA}).
Line profiles are more easily
resolved with HIRES, but gaps between the orders prevent simultaneous 
measurement of all emission lines of interest for accretion studies.

\section{Results}

\subsection{Radial and Rotational Velocities}

We determined $v\sin i$ values for the HIRES sample only -
the resolution of ESI is too low for any meaningful limits
given typical values for low mass young stars (e.g., Hartmann
et al. 1986; White \& Basri 2003, hereafter WB03).  The rotation velocities
were determined by cross-correlating photospheric absorption
features in small portions of each spectrum (see Tonry \& Davis 1979;
Hartmann et al. 1986).
We selected intervals of $\sim 20$ {\AA} in several different
orders containing relatively strong absorption features
with good S/N.  Values of $v\sin i$ were derived from the widths
of the cross-correlation peaks, which were calibrated using
template spectra of Gl 447 ($v\sin i < 2$ km s$^{-1}$, unresolved
at HIRES resolution; Delfosse et al. 1998) convolved with
a range of rotation velocities.  Results for the different orders
were then averaged together, with the standard deviation taken as
the final uncertainty.  We present these values in Table~\ref{hires_vel}.
The rotation velocities range from 8 - 35 km s$^{-1}$, with most
being $<$ 20 km s$^{-1}$.  The two fastest rotators, FP Tau and
LkCa 1, both at 34 km s$^{-1}$, show evidence of line doubling
in the Ca II triplet emission lines;
thus, the absorption line widths (which do not show obvious signs of
doubling) may be due to binarity rather than rapid rotation.

We determined heliocentric radial velocities through
similar cross-correlation techniques for both the HIRES and
ESI samples (see Table~\ref{hires_vel}).  For the HIRES spectra,
we used the dwarf standards Gl 402 and Gl 447 ($V_r = 
-2$ and  -31 km s$^{-1}$, respectively; Delfosse et al. 1998)
as templates.  For the ESI sample, we used
Gl 83.1 and Gl 406 ($V_r =  -28.567$ and 19.482 km s$^{-1}$,
respectively; Nidever et al. 2002).
Due to the relatively low resolution of the ESI data (R = 8,000),
the cross-correlation power is dominated by broad molecular features
which forced us to be selective in the wavelength regime and range
employed.  As for the $v\sin i$ measurements, results for
different wavelength ranges were averaged together, with the
errors taken from the standard deviation.

\subsection{Membership Information}

Many of the stars we have studied with HIRES and ESI
are candidate (as opposed to confirmed) members of
the Taurus, IC 348, Sigma Ori, Sco-Cen, and Ophiuchus stellar associations.
In these cases, membership is supported
by consistency in a color-magnitude diagram with the expectations
for a low-mass member at the distance and age of the cluster,
combined with a late spectral type derived from low dispersion spectroscopy.
Our higher dispersion data contain a variety of spectroscopic indicators
that may further inform membership status.  At the 1-10 Myr ages of
the associations contained in our sample, high lithium abundance
and low surface gravity are expected (we list Li equivalent widths
in Table~\ref{hires_vel}).  Further, we have checked
consistency of measured radial velocities
with those previously published for known cluster members.

In Taurus, we find that CIDA-13 lacks substantial Li I, although
its radial velocity is consistent with the average for Taurus members;
it is likely a foreground dwarf star.
The Sigma Ori cluster has two candidate members (SOri-45 and SOri-46)
with K I lines that are more similar to those of M dwarfs than of
low-gravity pre-main sequence stars, as well as radial velocities
which are substantially different from the other
Sigma Ori stars.  In the Upper Sco association, we find three stars
(UScoCTIO-85, UScoCTIO-121, UScoCTIO-132) which both lack substantial
Li I and have dwarf-like K I absorption.

As suggested in the tablenotes to Table 1, we question the membership
of these 6 stars in their respective associations.
Many of these stars are among the weakest H$\alpha$
emitters (Table~\ref{emlines}), and the emission line characteristics
in general are more consistent with active main sequence dwarfs rather than
young TTSs/brown dwarfs.  We do not include these objects in any of
the subsequent statistics or analysis.

\subsection{Emission Lines}

Emission line measurements for our entire sample are listed in
Table~\ref{emlines}, including equivalent widths for a variety of
permitted and forbidden lines, as well as the \ha line widths at 10\%
of peak intensity (a good discriminator of accretion versus
chromospheric emission; Muzerolle et al. 2000a, WB03).
Measurements for two main sequence late-M dwarfs (Gl 406 and LHS 2351)
are also given as a comparison.  The emission line spectra of the VLM
young objects are very similar to those of higher-mass young stars,
including both CTTSs and ``weak line" TTSs (WTTSs).  Balmer and He I
emission are virtually ubiquitous, and are typically stronger than
in the main sequence dwarfs.  The equivalent widths are larger than
seen in higher-mass WTTSs, but this is probably a result of the
cooler photospheres, rather than intrinsically stronger emission.
Ca II triplet emission is detected
in nearly half of the sample, while forbidden emission such as [OI]
$\lambda$6300 is detected in only $\sim$25\% of the sample
(though this smaller fraction may be due in part to the drop-off
in S/N of many of the spectra around this wavelength).

Most of the emission line profiles are narrow ($\lesssim 200$
km s$^{-1}$) and symmetric, typical of WTTSs and dMe stars.
However, 13 objects show broader profiles at \ha (see Fig.~\ref{halpha_obs}),
including many with blueshifted asymmetries, and several with
blueshifted absorption components (e.g., FP Tau, MHO-6, IC348-415).
Such features are more reminiscent of CTTSs.  The shapes of the upper
Balmer lines generally mirror \ha, while somewhat narrower,
and blueshifted absorption components decrease in strength as
one moves up the series.  Broad emission components in the Ca II and He I lines
are seen only in the earliest-type object in the sample,
Lk\ha 358.  This star is also the only one
that exhibits a high-velocity component at [OI].  Overall, the most
active of the $0.05-0.2 \; M_{\odot}$
VLM young objects have emission characteristics similar
to the {\it least} active among the 
well-studied $0.5 \; M_{\odot}$ CTTSs.

\subsection{Accretion Signatures}

We interpret our data in the context of accretion disks and accretion-driven
outflows associated with these young VLM objects, following our previous work
on higher mass TTS.  Hot UV/optical continuum veiling of photospheric
absorption features, as well as narrow emission components in Ca II and
He I, provide clues to the density, temperature structure and
filling factor of the accretion shock; these parameters may differ
from those characteristic of the well-studied $0.5 \; M_{\odot}$ stars,
as $M_*/R_*$ varies.
Observations of broad Balmer lines provide detailed information on
the nature of accreting magnetospheres.
Forbidden emission lines such as [OI] and [SII], and blueshifted
absorption components in permitted lines, are typically associated with
accretion disk-driven outflows (e.g. Hartigan, Edwards, \& Ghandour
1995; Calvet 1997).
We use all of these features to identify objects which are likely
undergoing active accretion, but focus primarily on veiling and
\ha emission line profiles in our analysis.

\subsubsection{Veiling}

One familiar manifestation of accretion in T Tauri stars is the veiling
of photospheric absorption features by excess continuum
emission produced in the accretion shock.  Veiling is typically defined
as the ratio of the excess flux to the photospheric
flux at a given wavelength: $r_{\lambda} = F_{\lambda,ex}/F_{\lambda,cont}$.
Consistent with the historical suggestions that accretion signatures
are weaker in lower mass stars, we find very little veiling in our VLM sample.
We have measured or placed upper limits on the veiling for each of
the 13 likely accretors (as defined from emission line indicators;
see below), using comparison spectra of similar
spectral type (e.g., Hartigan et al. 1989).  In order to avoid
mismatches due to gravity-sensitive features (see WB03
for a more detailed discussion), we opted to use the non-accreting
subsample, defined as those which clearly lack infalling gas signatures
at \ha, as veiling templates.  Since veiling at red optical wavelengths
is much less sensitive a diagnostic of accretion than \ha, these
templates should not have any intrinsic veiling of their own.
We note that WB03 measured an upper limit
$r \lesssim 0.22$ at 6500 {\AA} for one of our templates,
V410 X-ray 3, using a combination of dwarf and giant standard spectra.

Veiling measurements at several wavelength regions
are given in Tables~\ref{veiling1} and~\ref{veiling2}.
Due to the gaps in the wavelength coverage of the HIRES spectra,
we could not measure veiling at exactly the same wavelength regions
as ESI, hence we list these separately.  The wavelength
regions chosen for the ESI sample contain broad molecular absorption
features; we focused on these rather than narrow atomic features
in order to avoid line blending problems due to the lower resolution
of the ESI spectrograph.  For the two objects earlier than $\sim$M3,
Haro 6-28 and Lk\ha 358, the derived veiling is more uncertain
since the atomic features dominate.  These problems are much
less significant for the HIRES sample; however, the grid of
templates is less extensive, and slight mismatches due to differing
rotation rates result in greater uncertainties.  Note especially
the results for the V410 Anon 13 and CIDA 14 HIRES spectra -
the closest template in spectral type is about 0.5 subclasses
earlier, and rotates faster, which could result in slight
underestimates of the true veiling (although the results for
V410 Anon 13 are consistent with those of the ESI spectrum).

Most of the objects at M4 or later exhibit no significant
veiling (i.e., $\gtrsim 2 \sigma$ of the typical uncertainty
of $\pm 0.1$) at any of the wavelengths.  One low-mass object, IC348-415
does have significant veiling with $r \sim 0.5$ at 6200 {\AA},
though this value is highly uncertain
due to low S/N; this object also has the strongest \ha emission
of the entire sample, and our best profile model indicates
the highest $\mdot$ of the $>$M4 sample.  Nevertheless, we consider
this detection to be marginal at best.  Only in the 2 earliest-type
objects in the sample, Haro 6-28 and Lk\ha 358, do we measure robust
levels of veiling, indicating higher levels of accretion than
in the lowest-mass sample as a whole.  The veiling levels
are consistent with the general emission line characteristics
of the stars, such as the presence or absence of broad Ca II
triplet emission components, which were noted by Muzerolle, Hartmann,
\& Calvet (1998) to have a threshold behavior whereby
the broad component does not appear when the mass accretion rate
is $\lesssim 10^{-8} \; \msunyr$.

\subsubsection{\ha Profiles}

\begin{figure}
\plotone{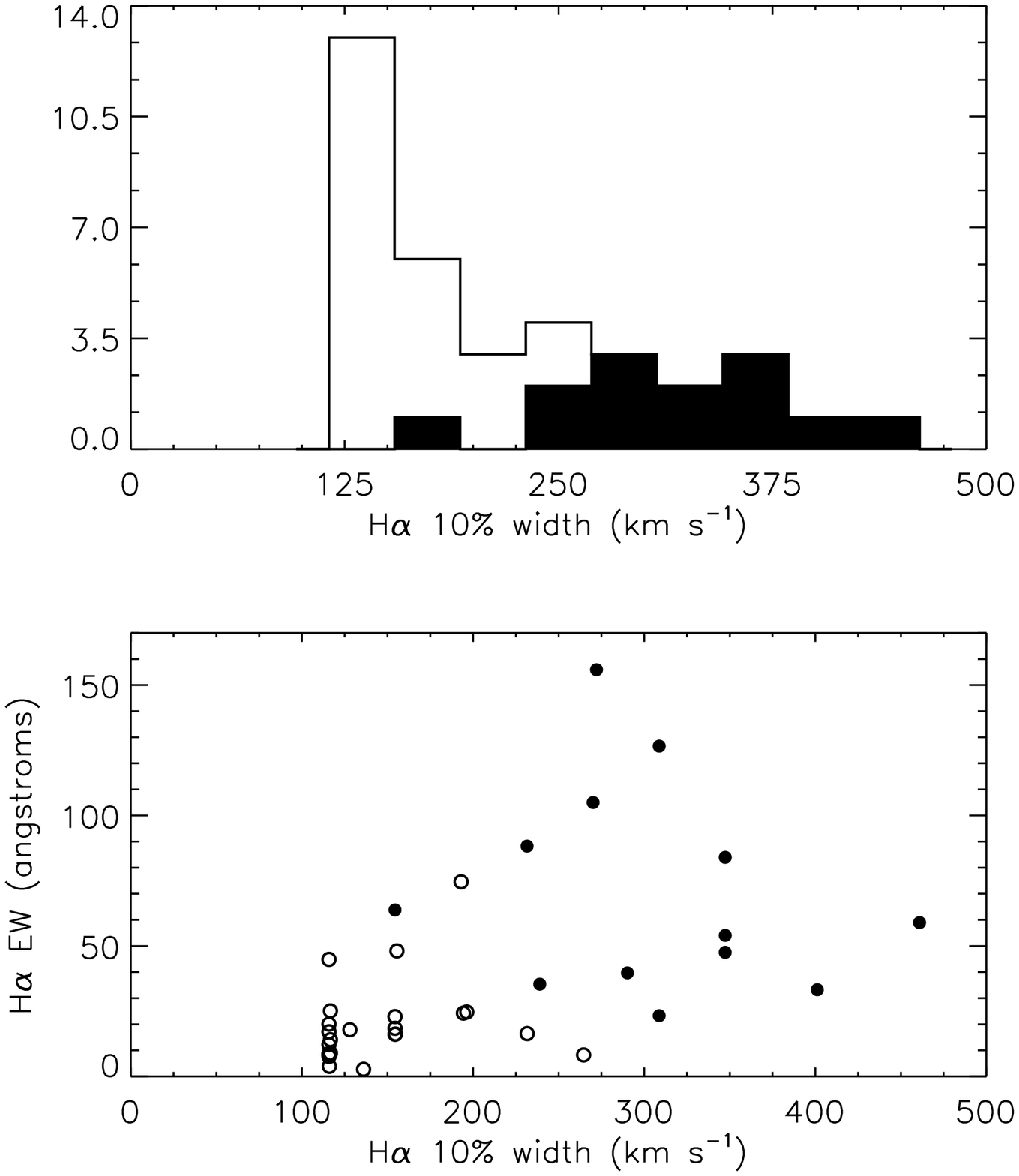}
\caption{Top panel: distribution of \ha 10\% line widths
for the entire sample (open histogram), and for the accretor
subsample (solid histogram).  Bottom panel: comparison of
\ha 10\% width versus equivalent width, for accretors
(solid circles) and non-accretors (open circles).
In both plots, objects where
the continuum at \ha was not detected were not included.
\label{widths}}
\end{figure}

Continuum-normalized \ha line profiles are shown in Figure~\ref{halpha_obs}.
Traditionally, the threshold
between CTTSs and WTTSs has been set at a canonical
boundary of \ha $EW = 10 \; \rm{\AA}$, which divides accreting from
non-accreting TTSs around the peak of the mass function in Taurus,
$\sim 0.5 \; M_\odot$.  However, this threshold
is inappropriate for much lower mass objects, where the decreased
optical photospheric continuum results in typically larger equivalent
widths despite comparable emission fluxes from chromospheric activity.
Recently, WB03 compared \ha 10\% line widths to
equivalent widths for TTSs of a wide range of masses, and calibrated
a spectral type-dependent equivalent width boundary between chromospheric
and accretion emission at \ha, down to late-M objects.
Figure~\ref{widths} shows the distribution of \ha 10\% line widths,
as well as a comparison with equivalent widths, for our {\it bona-fide}
young object sample.
Here, we also focus on the fact that the line profile shape is a much more
robust discriminator of chromospheric vs. accretion emission.
The broad ($>$200 km s$^{-1}$), asymmetric profiles produced in infalling
accretion flows are easily identified even in the lower-resolution
ESI spectra.  \ha emission from WTTSs, on the other hand,
is similar to that seen in chromospherically active main-sequence dwarfs
(e.g., Worden, Schneeberger, \& Giampapa 1981),
with relatively small equivalent widths and narrow ($<$200 km s$^{-1}$),
symmetric line profiles.  Our accretion models (see below) indicate that
accretion profiles in VLM objects should be detectable down to very low
accretion rates ($\sim 10^{-12} \; \msunyr$), much lower than
the sensitivity limits of any other optical diagnostics,
including continuum veiling\footnote{There is a possibility that
at least some of the broad \ha profiles
could be due to flares.  We believe this is unlikely for several
reasons.  Previously published profiles of M-dwarf flares sometimes show
a very weak broad gaussian component ($FWHM \sim 200$ km s$^{-1}$),
often redshifted, superposed on the (stronger) chromospheric profile,
and the equivalent widths are typically $\lesssim 20 \rm{\AA}$
(e.g., Abdul-Aziz et al. 1995; Abranin et al. 1998; Mart{\'\i}n 1999).
V927 Tau, which shows a similar profile, could be such a case.
In general, however, these flare profiles are not consistent
with our observations.
Also, our repeat observations of V410 Anon 13, separated by 9 months,
show a nearly identical \ha profile, indicating a relatively stable source
for the emission, thus favoring accretion over flare activity.}.

We have identified likely accretors on the basis of
line profile shape, namely whether the \ha lines are broad
($\gtrsim 200$ km s$^{-1}$) {\it and} asymmetric.  A total of 10 objects
meet both of these criteria, and we consider them to be strong candidates
for accretion.  Another 4 objects (V927 Tau, IC348-205, IC348-382,
and UScoCTIO-75) meet the line width criterion,
but are relatively symmetric.  Yet another 3 objects (MHO-4, MHO-5, and
UScoCTIO-128) meet neither of
the \ha accretion criteria, but show other potential signposts of accretion,
such as forbidden line emission or Ca II triplet emission fluxes stronger
than typical dwarf/WTTS levels.
Of these 7 objects with ambiguous accretion indicators, we believe
at least 3 are compelling cases for the presence of active accretion.
IC348-205 and IC348-382 have considerably stronger Balmer and Ca II
triplet emission than the VLM WTTSs, and, in both cases, \ha is quite broad,
if relatively symmetric, with two of the largest EWs in the entire sample.
MHO-5 exhibits relatively narrow and symmetric \ha emission; however,
the equivalent width is larger than typical of VLM WTTSs of similar
spectral type (e.g., WB03).  The profile shape of \ha in this object
is probably due to a pole-on geometry for the accretion flow, which
produces narrower, more symmetric emission profiles since most of
the high-velocity infalling gas is moving perpendicular to the line of sight
(\S 5.2; MCH).  MHO-5 also exhibits surprisingly strong [OI]
emission, which is never seen in higher-mass WTTSs.  Brice\~no et al. 1998
detected similarly strong \ha and [OI] emission in MHO-5, so we discount
variability due to flaring.  For our subsequent analysis,
we consider these 3 objects, plus the original 10 \ha-selected strong
candidates, as our total ``accretor" sample (see Tables~\ref{veiling1}
and~\ref{veiling2}).  The distribution of \ha 10\% line widths and equivalent
widths for this subsample are delineated in Figure~\ref{widths}.
The remaining 22 objects exhibit emission
characteristics consistent with chromospheric activity alone --
narrow, symmetric \ha, no forbidden emission, and little or no Ca II emission.

Yet another ambiguous case of interest is V927 Tau,
whose \ha profile includes a narrow, symmetric component with a small
self-absorption reversal at line center, all hallmarks of chromospheric
emission.  However, there is also a very weak, slightly asymmetric broad
component that could be due to accretion.  The equivalent width of the broad
component is roughly 2 {\AA}, more than an order of magnitude lower
than is typical of the accretor sample.  These characteristics are reminiscent
of flare activity, and we cannot reliably distinguish between that or
very low-level accretion without repeat observations, so we do not include
this object in the accretor sample.

\section{Modeling Accretion Flows}

\subsection{Continuum Excess Models}

Following the procedures of Calvet \& Gullbring (1998, CG98),
we calculate accretion shock emission and estimate the amount
of veiling expected for a given mass accretion rate, 
In short, infalling magnetospheric material accretes
onto the stellar photosphere through a shock just above
the photosphere. The material is assumed to be in plane-parallel
columns. In each column, soft X-ray emission from the shock heats
both the pre-shock region above the shock and the stellar photosphere
below it. The shock flux $F_{shock}$ is essentially the
sum of the spectra from the heated photosphere and the preshock
region. The spectral shape of the shock flux 
depends on the energy flux carried by each accreting column, $\curf$,
while the level of the emission depends on the surface
coverage of accreting columns, $f 4 \pi R_*^2$ (CG98).
The total luminosity of the shock is 
\begin{equation}
L_{shock} = f 4 \pi R_*^2 \curf,
\label{lshock}
\end{equation}
neglecting the intrinsic stellar flux, which is at least an order
of magnitude lower than the flux carried by the column in these cool stars.
At the same time, 
\begin{equation}
L_{shock} = \zeta L_{acc} = \zeta { {G M_* \mdot} \over  R_*},
\label{lacc}
\end{equation}
where $\zeta = 1 - 1/R_{mag}$, with $R_{mag}$ being the radius at
which magnetospheric infall begins.

We have calculated models for two generic stars with
parameters corresponding to our sample:
an M5 star 
($T_{eff} = 2900$ K, $M_* = 0.12 \, \msun$, $R_* = 1.2 \, \rsun$)
and an M6 star 
($T_{eff} = 2800$ K, $M_* = 0.05 \, \msun$, $R_* = 0.5 \, \rsun$).
Models were calculated for log $\curf$ = 10, 11, and 12, which
correspond to the range of energy fluxes that
provide better fits to the optical and ultraviolet 
excess fluxes in higher mass  T Tauri stars (CG98; Gullbring et al. 2000).
The emission from the shock was calculated
for $\mdot = 10^{-10} \; \msunyr, 10^{-9} \; \msunyr$ and
$10^{-8} \; \msunyr$, using eqs. (\ref{lshock}) and
(\ref{lacc}) to obtain the
filling factors $f$ for given  values of
$\mdot$ and $\curf$. We take $R_{mag} = 3$, in accord
with the \ha emission line models (\S 6).

To calculate the veiling, we compare the flux arising from the
accretion shock
with the intrinsic photospheric flux.
Here, we calculated the photospheric flux using bolometric
corrections from Luhman (1999) to convert from model luminosities
to absolute J magnitudes, and then used standard
colors to obtain corresponding V, R$_c$, and I$_c$ magnitudes.
Colors were taken from Leggett (1992) for types later than 
M4, and from Kenyon \& Hartmann (1995) for earlier types.
Fluxes at the effective wavelength of each filter
were obtained using the zero points of Bessell \& Brett (1988).
Fluxes were then scaled to the assumed model radii, and
then the veiling was calculated at each filter wavelength
by
\begin{equation}
r = 
{ F_{excess} \over F_{photosphere} } =
{ {f F_{shock}} \over {(1-f) F_{phot}}}.
\end{equation}

\begin{figure}
\plotone{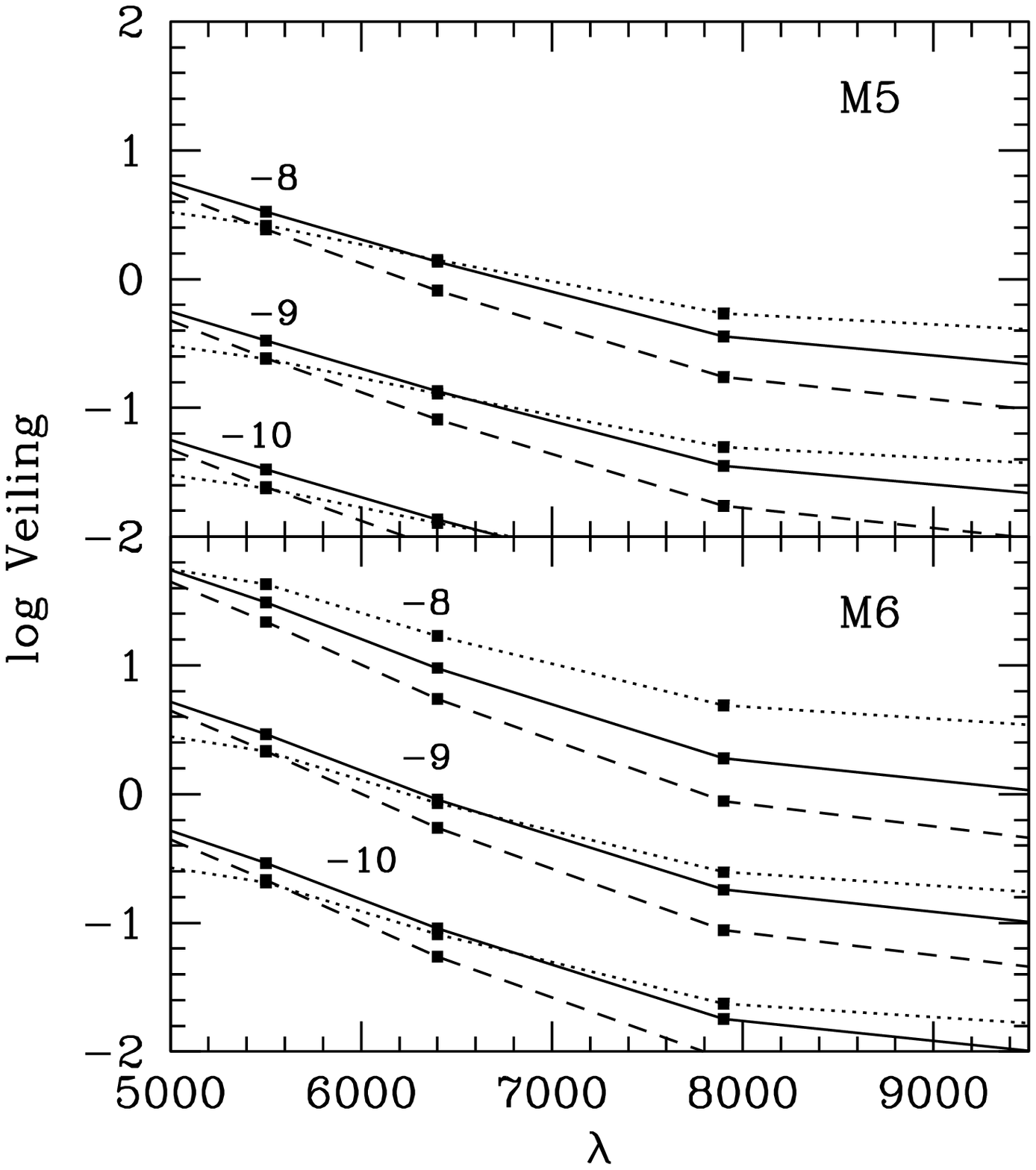}
\caption{Logarithm of the veiling as a function of
wavelength for an M5 star (upper panel) and an M6 star (lower panel),
representative of our sample.
Veiling values are shown for three values of the mass
accretion rate, log $\mdot$ = -10, -9, and -8, labeled 
accordingly. Shock models are calculated for
an energy flux 
log $\curf$ = 10 (dotted lines), 11 (solid lines), and
12 (dashed lines).
Photospheric colors are taken from Leggett (1992) and
Luhman (1999).  The horizontal dashed lines represent
the minimum measurable veiling level.
\label{veiling_models}}
\end{figure}

In Figure \ref{veiling_models}, we show the veiling
expected for the two model stars for 
log $\mdot$ = -10, -9, and -8, and log $\curf$ = 10, 11, and 12
(note the logarithmic scale). 
For a given $\mdot$, the veiling at 5500 {\AA} does not change
much with $\curf$, though the model with lower $\curf$ has a larger veiling
at longer wavelengths. This arises from the fact
that at optical wavelengths, the shock flux is dominated by 
quasi-blackbody emission from the heated photosphere,
which has a temperature $T_{hp} \sim ( \, 3\curf/ 4 \sigma \, )^{1/4}$ (CG98).
Thus, models with lower $\curf$ emit more at
redder wavelengths.

\begin{figure}
\plotone{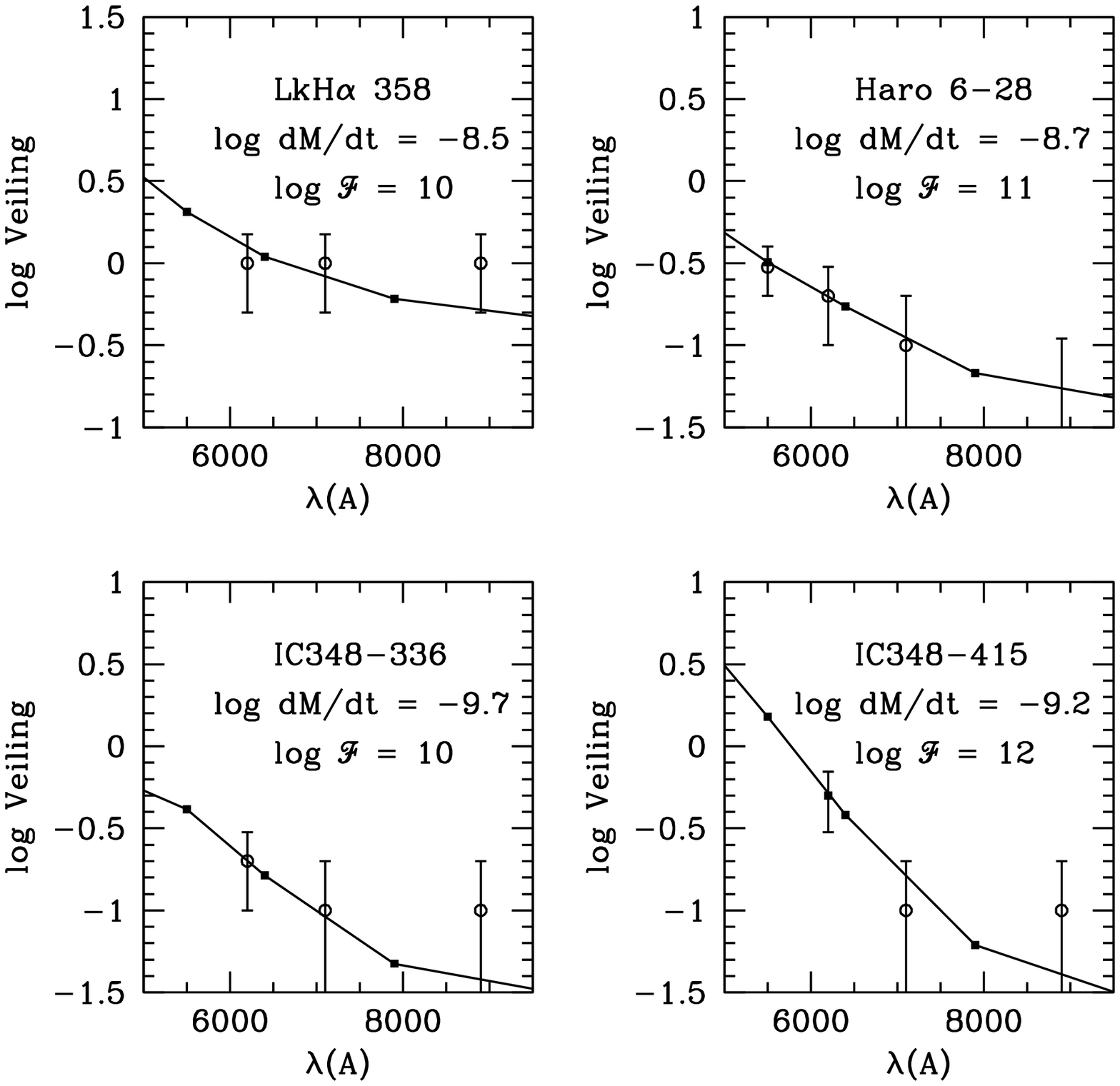}
\caption{Veiling measurements from accretors in Tables~\ref{veiling1}
and~\ref{veiling2} (open circles with error bars),
compared to shock models, cf. Figure \ref{veiling_models}
(solid squares, connected).
The value of $\curf$ indicated on each panel
matches the spectral shape of the veiling,
while $f$ is adjusted to match its level,
resulting in the indicated values of $\mdot$
in each case (with corresponding errors in log $\mdot$
of roughly $\pm$ 0.2-0.3).  The estimated filling factors
$f$ are 0.06 for Lk\ha 358, 0.002 for Haro 6-28, 0.01 for
IC348-336, and 0.0004 for IC348-415.
\label{veiling_objects}}
\end{figure}

It is apparent from Figure~\ref{veiling_models} that 
the negligible veiling measurements for most of the accretors
(Tables~\ref{veiling1} and~\ref{veiling2})
indicate $\mdot \lesssim 10^{-10} \; \msunyr$.
In Figure~\ref{veiling_objects}, we compare the detected
veiling for 4 accretors with predictions from shock models.
Energy fluxes that provide the best fit to the
observations, as well as the estimated mass accretion
rate are also shown.  The filling factors of the shock,
given in the caption of Figure~\ref{veiling_objects},
are rather small, $<< 1$\%, similar to the case of the more
massive CTTSs (CG98).  The accretion rates are in excellent agreement
with those determined from \ha models,
as described in the next section.

\subsection{\ha Emission Models}

Previously, we have developed radiative transfer models of
Balmer emission line profiles for $\sim 0.5 \; \msun$ CTTS
(MCH).  These models treat
line emission produced by magnetospheric accretion flows,
where gas accreting through a circumstellar disk is channeled
onto the star by the stellar magnetic field.  The model
assumes an ideal dipolar field geometry, and a specified
gas temperature distribution and mass accretion rate;
the gas density distribution is determined directly from
the geometry and the gas velocity (a function of the
mass and radius of the star).  Further details can be found
in MCH.  Typical results show
a characteristic profile shape with blue asymmetry, due to
occultation effects and some absorption of
the red wing where the receding part of the flow is
seen in projection against the star and accretion shock.
Frequently, outright redshifted absorption components are seen
in the upper Balmer lines, but rarely in \ha because
of thermalization and opacity broadening effects.
The qualitative similarity of most of the asymmetric \ha
profiles of the sample presented herein to these general
model characteristics lends strong support for our application
of these models to the lower-mass objects.

Here, we employ the same model, using two values of the
stellar mass (0.05 and 0.15 $\msun$) and radius (0.5 and 1 $\rsun$)
that span the range appropriate
for our VLM sample (see Table~\ref{info}).  The mass range will change
somewhat depending on which evolutionary tracks are used; however,
this should not affect our modeling applications since the model
gas velocities, and, hence, line widths, are only sensitive to
$\sqrt{M}$, and the derived masses are unlikely to be off by more
than a factor of two.  The radius is dominated by uncertainties
in the spectral type - $T_{eff}$ transformation; a 200 K error
in $T_{eff}$ translates to $\sim 30 \%$ uncertainty in the radius.
Since the model gas density does not depend very strongly on
the radius (see MCH, and references therein), our model constraints
are insensitive to this level of uncertainty.

\begin{figure}
\plotone{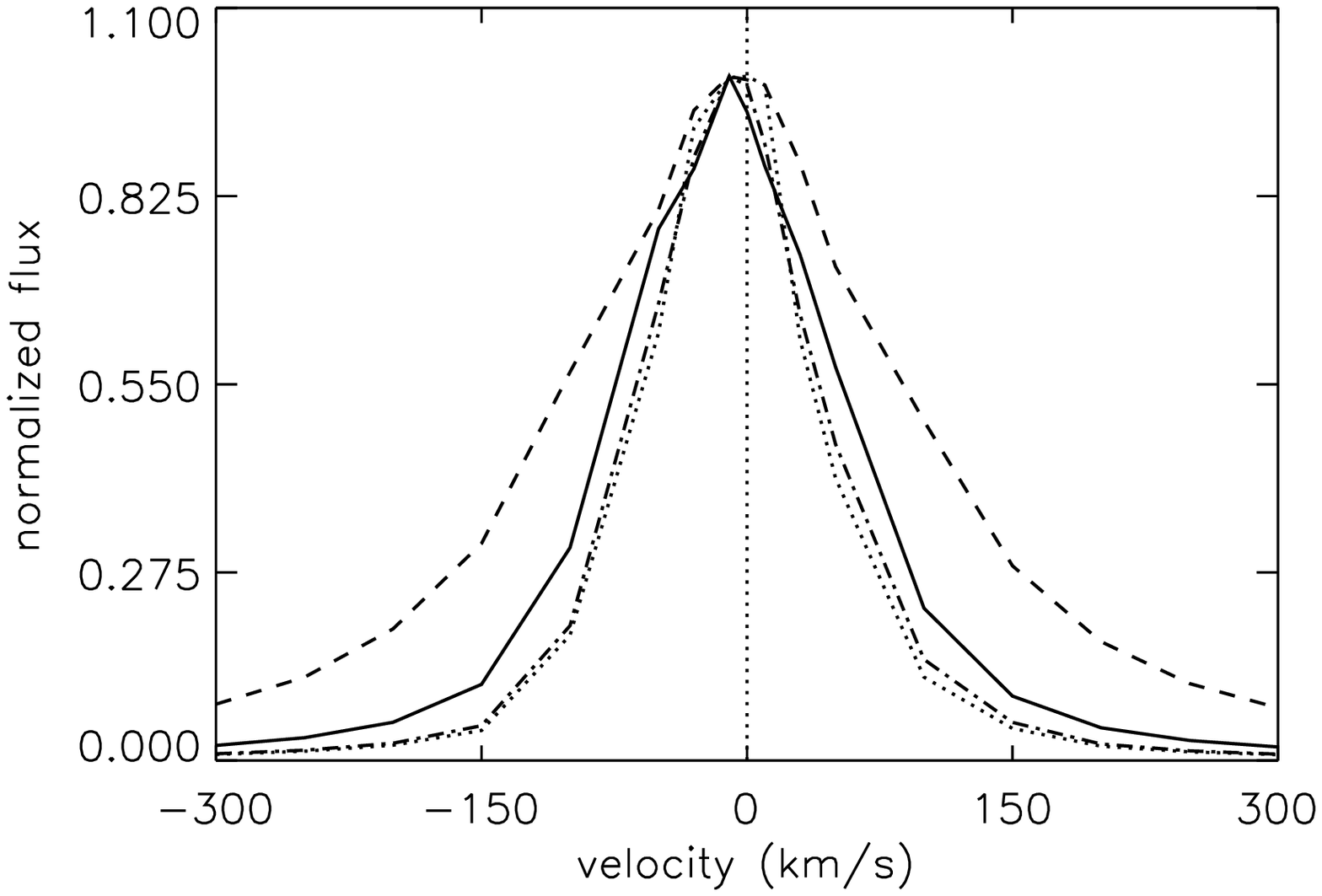}
\caption{Model \ha profiles, as a function of magnetospheric
density and temperature.  All models calculated with
$M_*=0.05 \, \msun$, $R_*=0.5 \, R_{\odot}$, $i=60^\circ$,
and $R_{mag}=2.2-3 \; R_*$.  Solid line: $\mdot=10^{-9} \,
\msunyr$, $T_{max}=8000 \; K$.  Dashed line: $\mdot=10^{-9} \,
\msunyr$, $T_{max}=10,000 \; K$.  Dot-dashed line:
$\mdot=10^{-10} \, \msunyr$, $T_{max}=10,000 \; K$.
Dotted line: $\mdot=10^{-10} \, \msunyr$, $T_{max}=12,000 \; K$.
\label{model_grid}}
\end{figure}

We use the same temperature
constraints as those found to reproduce the 0.5 $\msun$ CTTS
profiles, which should be a reasonable assumption since
the gas heating is likely mechanical and independent of the
stellar $T_{eff}$.  These constraints result in an inverse
proportionality between the gas temperature and density.
Since the accretion rates of all the objects we are considering
here are $\lesssim 10^{-9} \; \msunyr$ (as inferred from the lack
of continuum veiling), the accretion flow
gas temperatures must be $\gtrsim$ 10,000 K to obtain sufficient
\ha emission, thus the gas is
almost completely ionized.  As an independent check of the adopted
temperature constraints, we show several \ha models as a function
of both temperature and $\mdot$ in Figure~\ref{model_grid}.
The models at the higher value of $\mdot$ ($10^{-9} \; \msunyr$)
exhibit broader line widths than those at the lower $\mdot$,
as opacity broadening effects kick in due to the higher
density; the profile is still broader at a lower temperature
of $T_{max}=8000$ K (temperatures lower than this are not
applicable at this value of $\mdot$ as the line flux becomes
much lower than observed).  At the lower value of $\mdot$
($10^{-10} \; \msunyr$),  the density is too low to produce
opacity broadening in the wings, and the profile shape is nearly identical
for $T_{max}=10,000$ and 12,000 K.
Thus, the model temperature can be constrained enough
to obtain more accurate constraints on the mass accretion rate.

As for the flow geometry, we have selected the fiducial
magnetosphere inner and outer radius values from MCH
(2.2-3 R$_*$).  A larger magnetosphere would produce higher line
fluxes and more strongly centrally-peaked profiles for a given
accretion rate, while a smaller magnetosphere would
produce redward asymmetries in the line profiles due to
stellar occultation effects.  The fiducial size is also
consistent with inner hole sizes inferred from observed
($H-K$) and ($K-L$) excesses (Table~\ref{info}; Meyer et al.
1997; Liu et al. 2002).  Varying the magnetospheric size
by roughly 20\% may still produce an adequate match to the observed
profiles, and still be consistent with infrared excesses.
This results in an uncertainty in our $\mdot$ constraints of
about a factor of 2-3 (see MCH for a detailed comparison of
magnetospheric size, accretion rate, and line flux).

Finally, we have not included the effects
of rotation in the models presented here.  MCH showed that
for $V_{rot} \lesssim 20$ km s$^{-1}$, consistent with most
of our $v\sin i$ measurements (\S 4.4), rotation does
not have a significant effect on the line profile.
A few of the profiles we model below show evidence for
faster rotation, but since we cannot measure $v\sin i$,
we have elected not to include rotation effects for these.
$\mdot$ constraints are not affected by this omission,
since the model gas density is virtually unchanged for
the plausible range of rotation rates in the low mass objects.

\begin{figure}
\plotone{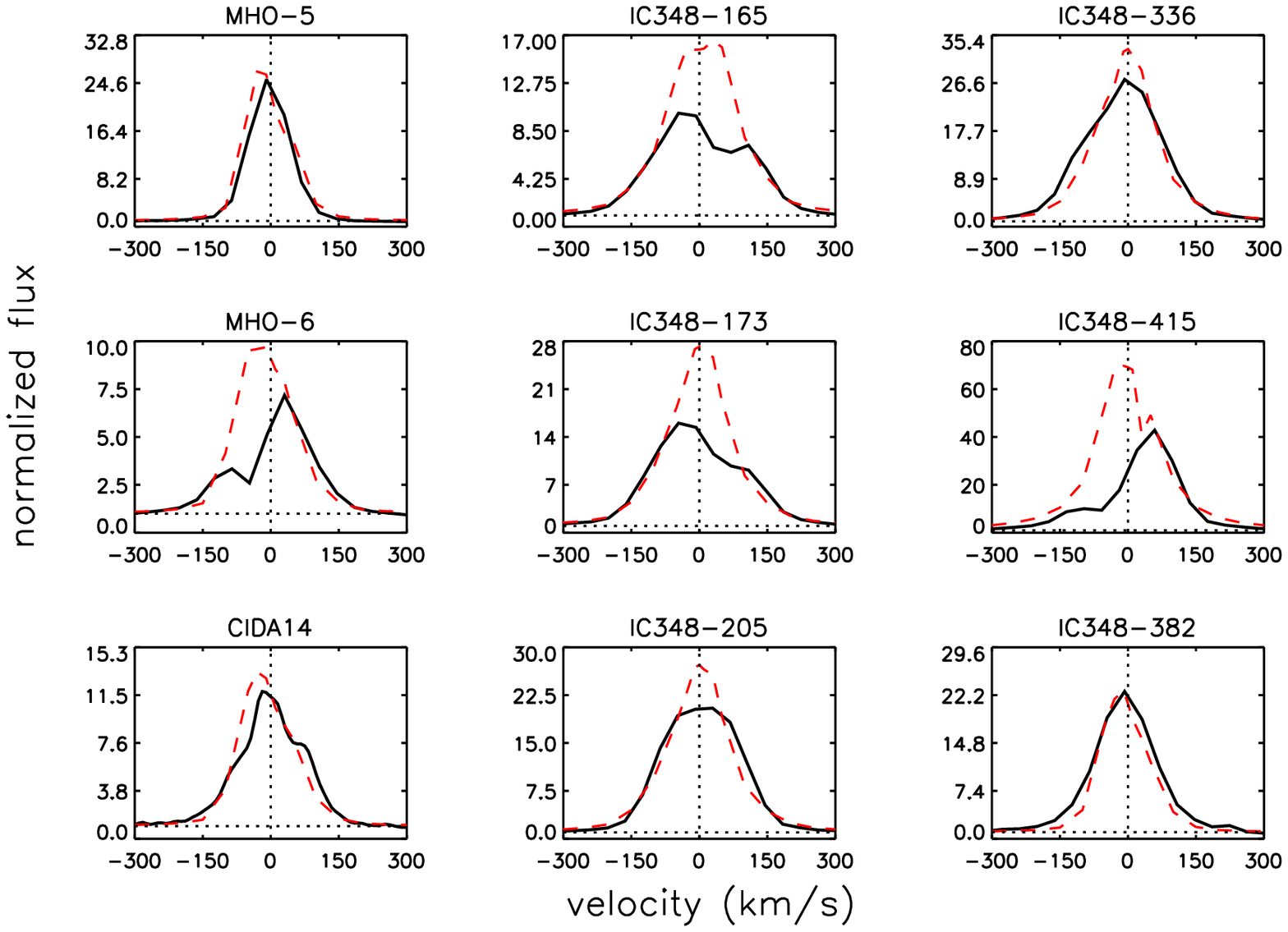}
\caption{Comparison of observed and model \ha profiles for the 9
accretors with spectral types $>$ M4.  Note that the models
do not include a wind component, and hence cannot reproduce
the blueshifted absorption components seen in
MHO-6 and IC348-415.  Discrepancies near line center 
are likely due to rotation effects or a breakdown
of the Sobolev approximation (see text).
\label{ha_mod}}
\end{figure}

For each accretor \ha profile, we constructed a model by adopting
the above parameter constraints, varying only the mass accretion
rate and inclination angle to find the best match to the observed
profile shape and flux.  A range of accretion rates were explored,
$10^{-12} < \mdot < 10^{-9} \; \msunyr$, with the lower limit
set where accretion emission is produced at an undetectable level,
and the upper limit set by the veiling constraints.
The inclination angle was varied from pole-on to edge-on
orientations of the star/disk system.
Our best model matches are shown in Figure~\ref{ha_mod},
with parameters listed in Table~\ref{model_param}.
We have not modeled the 3 accretors earlier than M5,
due to the temperature parameter degeneracy problem mentioned above
(we did derive accretion rates from the veiling for two of these,
Haro 6-28 and Lk\ha 358, with $\mdot > 10^{-9} \; \msunyr$ in both cases;
see Figure~\ref{veiling_objects}).
We also do not show a model for V410 Anon 13, the subject of an earlier
paper (Muzerolle et al. 2000b).  Note that in two objects
shown in Figure~\ref{ha_mod}, MHO-6 and IC348-415, the \ha profiles
show blueshifted absorption components.  Such features are
produced in accretion-powered winds, which are not treated
in the model calculations.  In these cases,
we have tried to match only the red and high-velocity
blue wings.

The overall agreement between the observed and model profiles
is remarkably good, especially considering the idealized
model geometry.  In reality, the accretion flows are probably
not axisymmetric, and in many cases may be broken up into
discrete ``streams" (e.g., Muzerolle et al. 2000b).
The discrepancies near line center for some
of the IC 348 objects (higher predicted flux than observed)
may be a result of neglecting rotation
in the model calculations; we do not have rotation velocities
for those objects, which could conceivably be rotating
much faster than typical of the Taurus objects in the HIRES subsample.
Previous models we have generated for higher-mass stars
with rapid rotation generally produce somewhat broader,
less-strongly peaked profiles with doubling at line center.
In addition, the assumption of a large velocity gradient
implicit in our use of the Sobolev approximation in the model
calculations for the line source function break down
near line center, which mainly probes
the outermost region of the magnetospheric flow where velocities
are relatively small and densities are correspondingly large.
Thus, we may be overestimating the amount of flux that can escape from
the outer regions of the magnetosphere.
Nevertheless, these models  provide quantitative estimates of
mass accretion rates (which we show in Tables~\ref{veiling1}
and~\ref{veiling2}) that are
accurate to within a factor of $\sim$3-5, given the preceding caveats.
Note that these \ha-derived accretion rates are consistent
with the values and upper limits derived from the veiling,
as shown in the previous section.
Although our $\mdot$ estimates are not as accurate as UV
excess measurements, they are likely the best achievable
constraints for VLM objects at present, given the generally small
accretion rates, which yield only a small amount of excess emission
in the UV.

\section{Discussion}

The clear signatures of accretion exhibited in roughly 30\%
of our late-type sample indirectly reveal the presence of
circumstellar accretion disks, and indicate that such disks
are relatively common in young VLM objects.
Observations of infrared excesses yield further evidence
of disk accretion,
and several recent studies along these lines have shown
that the fraction of VLM young stars and brown dwarfs
with disks is indeed -- and not surprisingly --
similar to that of the higher mass TTSs
(e.g., Muench et al. 2001; Liu, Najita \& Tokunaga 2002).  
Examination of Table~\ref{info} shows that
a fairly large fraction of our sample exhibits
near-infrared excess, indicative of circumstellar disks.
The level of excess at these wavelengths suggests circumstellar disks
with fairly small inner holes, relatively consistent with
the inner magnetospheric radius of our \ha models.
However, most of the excesses are fairly small ($\Delta(H-K) \lesssim 0.2$),
and we cannot definitively rule out spots, spectral type mismatches,
or cooler, unresolved companions as possible sources of apparent excess.

In any case, the general agreement between the observed and model \ha profiles
we present here shows that magnetically mediated accretion from a disk
is applicable to VLM young objects.
Thus, the disk accretion paradigm, which is a common (if not
universal) aspect of early stellar evolution,  appears to work
over a very large range of masses, from substellar objects
with $M<0.08 \msun$ to at least the highest-mass TTS at $M \sim 2 \; \msun$
(and possibly to even higher masses, including the Herbig Ae/Be
stars, some of which exhibit magnetospheric infall signatures;
Muzerolle et al. in preparation).  A key parameter in
the physical description of accretion
is the mass accretion rate, which guides both the growth of
a young star/brown dwarf and the evolution of its disk.
Mass accretion rates have been measured extensively at typical
TTS masses of $0.5-1.5 \; \msun$ (e.g. Valenti et al. 1993;
Hartigan, Edwards, \& Ghandour 1995; Gullbring et al. 1998;
Hartmann et al. 1998; White \& Ghez 2001); however, in most cases
the mass ranges probed have not been sufficiently broad or
complete to effectively study any mass dependences.  Rebull et
al. (2000; 2002) did estimate $\mdot$ for $M \sim 0.25 - 1 \; \msun$
stars in Orion and NGC 2264 using $U$-band excesses, and found
a loose correlation with $\mdot \propto M$.  However, their
photometrically-selected samples are incomplete at average or
lower values of $\mdot$, and rather large extinction uncertainties
result in a more uncertain $U$-excess calibration.

More recently, WB03 attempted to measure accretion
rates from continuum veiling in a sample of VLM objects in
Taurus (including four objects in common with our sample).
Most of their sample lacked measurable veiling (as does ours),
yielding upper limits on accretion rates of the order $10^{-10} \; \msunyr$.
They did find two objects
\footnote{Both of these, GM Tau and CIDA-1,
were previously identified from low-resolution spectra as
continuum objects, with indeterminate spectral types.}
with $M<0.2 \; \msun$, including
one probable brown dwarf, that exhibited considerable veiling,
yielding accretion rates of approximately $2-3 \times 10^{-9} \;
\msunyr$.
Together with our previously reported accretion rate for VLM
object V410 Anonymous 13 (Muzerolle et al. 2000b), and values
for higher-mass Taurus TTS, WB03 also find an
approximate correlation of $\mdot \propto M$.  However, the number
of $\mdot$ values at the lowest masses that are not upper limits
is small.

\begin{figure}
\plotone{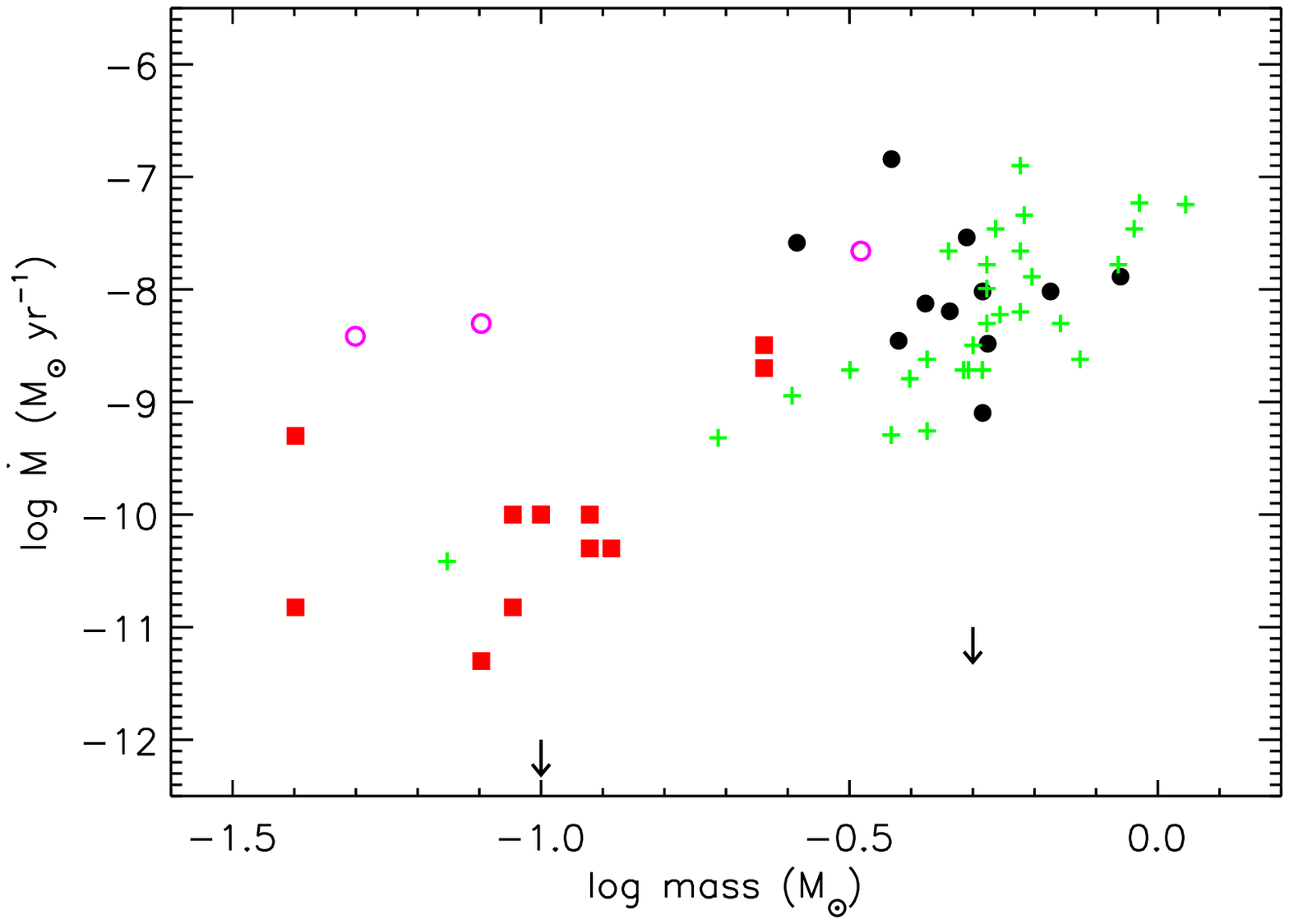}
\caption{Mass accretion rates as a function of stellar/substellar mass.
Squares represent $\mdot$ values determined from \ha modeling and veiling
in this work and Muzerolle et al. (2000a); solid circles are from
Gullbring et al. (1998) (omitting binaries); pluses are from
White \& Ghez (2000) (including binary components); open circles are
from White \& Basri (2003).  The arrows represent lower limits to
$\mdot$ at 0.1 and 0.5 $\msun$ measurable from \ha, as determined
from the accretion models.
\label{mass_mdot}}
\end{figure}

With \ha modeling, we can probe to accretion rates several orders
of magnitude lower than possible from veiling/UV excesses.
Figure~\ref{mass_mdot} shows our derived $\mdot$ values as
a function of mass, along with the WB03
results, and values for higher-mass TTS from
Gullbring et al. (1998) and White \& Ghez (2001).
Note that WB03 and White \& Ghez (2001) used the Baraffe et al.
(1998) pre-main sequence evolution tracks to determine masses;
we have scaled those masses, as well as their values of $\mdot$
(since they used $\mdot = L_{acc} R_*/GM_*$), to the
D'Antona \& Mazzitelli (1998) evolution tracks in order to be
consistent with our results.
There is a clear correlation between mass accretion rate
and stellar/substellar mass, with an approximate proportionality
of $\mdot \propto M^2$.  Such a dependence is steeper than
in previous investigations.

There are several
caveats that must be emphasized when interpreting the above result.
One is that the derived (sub)stellar masses are dependent on which
evolutionary tracks are used; for example, the Baraffe et al. (1998)
tracks yield systematically larger masses, by a factor of $\sim$ 1.5-2.
This also affects accretion rates derived from veiling or the UV
excess, since those methods require the stellar mass and radius
in order to derive $\mdot$ from the accretion luminosity.
However, we checked the mass-$\mdot$ comparison
using Baraffe et al. (1998) masses, and find the same proportionality.
 
Another caveat is that objects accreting at levels low enough that \ha emission
from the accretion flow (or UV emission from the accretion
shock) is undetectable, would be missed in all surveys conducted
to date.  Such an effect is probably more relevant
to the VLM objects since the minimum $\mdot$ at $0.1 \; \msun$
needed to produce observable \ha emission, as determined from
our models, is within about a factor of 2 of the lowest measured
value in our VLM sample.  At $0.5 \; \msun$, the models
indicate a higher minimum $\mdot$ (since the higher mass of
the central object results in higher infall velocities, which
in turn results in smaller magnetospheric gas densities for
the same accretion rate), but one which is over
one order of magnitude lower than the lowest measured value at
that mass.  Thus, at least at the higher-mass TTSs, the observed
lower limit to $\mdot$ is probably real\footnote{This is also
somewhat dependent on age, since older objects at the same mass
have smaller radii, which results in somewhat higher densities
at the same $\mdot$.  Nevertheless, all of the objects plotted in
Figure~\ref{mass_mdot} are at roughly the same age, 1-3 Myr.
See Muzerolle et al. (2000a) for results at 10 Myr.}.

A final caveat is that VLM objects accreting at rates higher than
those measured thus far could be missed due to either
veiling completely filling in photospheric features, or
very high extinctions.  However, the number of so-called
``continuum stars" is fairly small (6 in Taurus, as listed in
Kenyon \& Hartmann 1995), and not necessarily all are VLM objects.
One of three Taurus continuum stars observed by WB03,
GN Tau, turned out to be M2.5; the other three in Hartmann \&
Kenyon (1995) have bolometric luminosities higher than likely
for accreting VLM objects.  Also, there is no evidence
for any correlation between extinction and accretion activity
in TTSs (not including embedded ``Class I" objects).
Furthermore, in the extinction-limited sample of Brice\~no et al.
(2002), no correlation between extinction and (sub)stellar mass
up to $A_V \sim 4$ was seen, thus there is no evidence that young
brown dwarfs are selectively more extincted than higher-mass TTSs.

The physical basis for a mass vs. $\mdot$ relation is
not entirely clear.  In the standard model of viscous
disk accretion (e.g., Lynden-Bell \& Pringle 1974),
the disk mass accretion rate is proportional
to the disk surface density times the viscosity.
The currently favored viscosity mechanism, the Balbus-Hawley
instability (e.g., Balbus \& Hawley 1998), requires sufficient
ionization of the disk gas to operate efficiently.
A likely source of ionization of T Tauri disks is X-ray radiation
produced by magnetic activity (Glassgold, Najita, \& Igea 1997;
Igea \& Glassgold 1999).
Interestingly, recent studies have revealed a fairly strong
correlation between X-ray luminosity and mass in young stellar objects
(Feigelson et al. 2003; Flaccomio et al. 2003; Mokler \& Stelzer 2002;
Preibisch \& Zinnecker 2002), where $L_X$ on average
decreases by a factor of $\sim 50-100$ from 1 to 0.1 M$_{\odot}$.
It is possible that the lower X-ray flux produced by less massive
young stars could result in smaller accretion rates by way of
decreased disk ionization fractions, driving the correlation we find
between $\mdot$ and mass.
However, it is not clear whether the difference in $L_X$
over the mass range represented in Figure~\ref{mass_mdot} can create
enough of a change in disk ionization levels to produce a
relation as steep as $\mdot \propto M^2$.  On the other hand,
the correlation we have determined may be artificially steep,
due to the possible selection effects described above.
In any case, our results should provide interesting
new motivation for continued theoretical work on disk
accretion and evolution in young stars and VLM objects.

It should be noted that the mass-$\mdot$ relation we find is for
objects with a relatively narrow range of ages, $\sim 1-3$ Myr.
Thus, we cannot make any conclusions about whether such a relation
holds at younger ages, i.e., at the embedded proto(sub)stellar phase.
Most of the mass is probably accreted during the first few $\times
10^5$ years, either in the initial cloud collapse, or via FU Orionis-
type outbursts, which may be different from typical
viscous accretion and hence may be unrelated to the scenario we suggest
in the previous paragraph.  The accretion rates measured here
are indicative of a more quiescent, steady accretion phase, and do not
have much impact on the subsequent mass evolution of these objects.

We can comment on the fraction of accretors as a function of
(sub)stellar age and mass.  We have found 13 low-mass accretors
with $\mdot > 10^{-12} \; \msunyr$,
all of which reside in the young (1-3 Myr) IC 348 or Taurus associations.
Conversely, we have found no low mass accretors 
in the somewhat older (3-8 Myr) Sigma Ori or Sco Cen associations. 
Consistent with previous results on the evolution of disk accretion 
in higher mass, $\sim0.5 \; M_\odot$, 
stars (e.g. Muzerolle et al 2000a; Muzerolle et al. 2001b), we find here
an accretion timescale on the order of several Myr 
for objects near the substellar boundary.
At mass $\sim0.5 \; M_\odot$ and age $\sim 2$ Myr,
the fraction of stars identified as accretors is about
50\% (e.g., Kenyon \& Hartmann 1995 for Taurus).
Here, although we have found only about 30\% of our total $<0.2 \; M_\odot$ 
sample to be accreting, $\sim$50\% of the Taurus sample is accreting,
albeit at lower rates.

\acknowledgements

We thank Russel White, Michael Meyer, and Kevin Luhman
for helpful discussions and advice, and Suzan Edwards for
a critical reading of the manuscript.
This work was supported in part by NASA Origins of
Solar Systems grant NAG5-9670 to NC and LH.  CB received
partial support from grant S1-2001001144 of the Fondo
Nacional de Ciencia y Tecnolog{\'\i}a (FONACYT) of Venezuela.

\begin{deluxetable}{lcccccccccc}
\tabletypesize{\small}
\rotate
\tablewidth{0pt}
\tablecaption{(Sub)Stellar Properties of the Observed Sample
\label{info}}
\tablehead{
\colhead{object} & \colhead{Spec. Type} & \colhead{log $L_{bol}$ (L$_{\odot}$)}
& \colhead{$I$} & \colhead{$J-H$} & \colhead{$H-K$} & \colhead{$K$} & \colhead{$A_V$}
& \colhead{$\Delta(H-K)$} & \colhead{log age (yr)} & \colhead{mass (M$_{\odot}$)}}
\startdata
     CIDA 13  & M3.5   & -1.47   &13.95  &  0.61  &  0.20  & 11.85 &   0.35&   -0.08 & 6.94&0.19  \\                   
     CIDA 14  & M5     & -0.64   &12.24  &  0.67  &  0.33  &  9.41 &   0.34&   -0.01 & 5.10&0.12      \\               
       FN Tau  & M5     & -0.30   &\nodata&  0.95  &  0.50  &  8.25 &   1.35&    0.10 &$<$5.00&0.11        \\             
       FP Tau  & M2.5   & -0.49   &11.33  &  0.69  &  0.25  &  8.97 &   0.24&    0.00 & 5.88&0.22              \\       
    Haro 6-28  & M2.5   & -0.82   &\nodata&  1.07  &  0.74  &  9.27 &   1.77&    0.39 & 6.23&0.23                  \\   
       LkCa 1  & M4     & -0.44   &11.07  &  0.77  &  0.24  &  8.69 &   0.00&    0.01 & 5.79&0.21                     \\
     LkHa 358  & K7-M0  &  0.88   &16.05  &  2.09  &  1.21  &  9.69 &  13.6&     0.19 &$<$5.00&0.23                     \\
       MHO-4  & M7     & -1.14   &14.32  & \nodata& \nodata&\nodata&   0.97&\nodata & 5.06&0.06                     \\
       MHO-5  & M6     & -1.18   &13.72  &  0.63  &  0.48  & 10.05 &   0.11&    0.11 & 5.63&0.09                     \\
       MHO-6  & M4.75  & -1.16   &13.80  &  0.71  &  0.42  & 10.63 &   0.86&    0.06 & 6.31&0.13                     \\
       MHO-7  & M5.25  & -0.97   &13.18  &  0.65  &  0.29  & 10.15 &   0.40&   -0.06 & 5.62&0.12                     \\
       MHO-8  & M6     & -1.02   &13.63  &  0.74  &  0.38  &  9.74 &   0.62&   -0.02 & 5.42&0.09                     \\
       MHO-9  & M5     & -0.47   &12.95  &\nodata & \nodata&\nodata&   2.22&\nodata  &$<$5.00&0.11                     \\
 V410Anon 13  & M5.75  & -1.45   &16.59  &  1.24  &  0.71  & 10.95 &   3.83  &  0.12 & 6.51&0.08                     \\
  V410 Xray-3  & M6     & -1.20   &14.18  &  0.71  &  0.39  & 10.39 &   0.80  & -0.02 & 5.66&0.09                     \\
 V410 Xray-5a  & M5.5   & -1.29   &15.37  &  1.24  &  0.61  & 10.14 &   2.57  &  0.11 & 6.33&0.10                     \\
     V927 Tau  & M3.5   & -0.46   &11.46  &  0.86  &  0.31  &  8.68 &   0.38  &  0.02 & 5.17&0.15                     \\
   IC348-165  & M5.25  & -0.98   &16.07  &  0.81  &  0.50  & 11.77 &   2.41  &  0.02 & 5.64&0.12                     \\
   IC348-173  & M5.75  & -0.95 &\nodata  &  0.72  &  0.53  & 11.85 &   1.42  &  0.09 & 5.40&0.10                     \\
   IC348-205  & M6     & -1.11 &\nodata  &  0.69  &  0.65  & 12.15 &   1.21  &  0.21 & 5.52&0.09                     \\
   IC348-336  & M5.5   & -1.40  & 17.61  &  0.91  &  0.63  & 13.27 &   3.12  &  0.10 & 6.42&0.10                     \\
   IC348-363  & M8     & -2.12  & 17.95  &  0.64  &  0.69  & 13.50 &   0.00  &  0.24 & 5.37&0.02                     \\
   IC348-382  & M6.5   & -1.91  & 18.92  &  0.87  &  0.76  & 13.72 &   2.77  &  0.20 & 5.69&0.04                     \\
   IC348-407  & M7     & -2.01  & 19.71  &  0.88  &  1.41  & 13.97 &   3.55  &  0.78 & 5.37&0.03                     \\
   IC348-415  & M6.25  & -2.00  & 18.23  &  0.67  &  0.12  & 14.03 &   1.35  & -0.34 & 6.07&0.04                     \\
   IC348-454  & M5.75  & -2.19  & 17.82  &  0.88  &  0.31  & 14.31 &   0.11  & -0.04 & 6.84&0.05                     \\
   IC348-478  & M6.25  & -1.92  & 18.58  &  1.13  &  0.40  & 14.64 &   2.27  & -0.11 & 5.98&0.04                     \\
       SOri-12&   M6   &   -1.51&   16.47&    0.56&    0.36&   13.28 &   0.00&    0.00&  6.31&0.07                     \\
       SOri-17&   M6   &   -1.70&   16.95&    0.58&    0.40&   13.79 &   0.00&    0.04&  6.46&0.06                     \\
       SOri-25&   M6.5 &   -1.75&   17.16&    0.51&    0.40&   13.76 &   0.00&    0.02&  5.62&0.04                     \\
     SOri-29  & M6     & -1.81  & 17.23  &  0.53  &  0.34  & 13.96   & 0.00  & -0.02  &6.54&0.05     \\
       SOri-40&   M7   &   -2.09&   18.09&    0.54&    0.36&   14.59 &   0.00&   -0.05&  5.39&0.02       \\              
       SOri-45&   M8.5 &   -2.62&   19.59&    0.65&    0.37&   15.66 &   0.00&   -0.09&  5.98&0.02           \\          
     SOri-46  & M8.5   & -2.71  & 19.82  &  0.65  &  0.37  & 15.66   & 0.00&   -0.09  &6.09& 0.02                \\     
  UScoCTIO-66  & M6     & -1.55  & 14.85  &  0.60  &  0.38  & 11.92   & 0.00&    0.02  &6.34&0.07                     \\
  UScoCTIO-75  & M6     & -1.64  & 15.08  &\nodata &\nodata &\nodata  & 0.00&\nodata   &6.41&0.06                     \\
  UScoCTIO-85  & M6     & -1.70  & 15.23  &  0.58  &  0.39  & 11.95   & 0.00&    0.03  &6.46&0.06                     \\
 UScoCTIO-100  & M7     & -1.78  & 15.62  &  0.62  &  0.40  & 11.83   & 0.00&   -0.01  &5.31&0.03                     \\
 UScoCTIO-109  & M6     & -2.03  & 16.06  &\nodata &\nodata & \nodata & 0.00&\nodata   &6.71&0.04                     \\
 UScoCTIO-121  & M6     & -2.19  & 16.46  &  0.61  & \nodata& \nodata & 0.00&\nodata   &6.79&0.04                     \\
 UScoCTIO-128  & M7     & -2.37  & 17.09  &  0.63  & \nodata& \nodata & 0.00&\nodata   &5.47&0.02                     \\
 UScoCTIO-132  & M7     & -2.59  & 17.63  &  0.77  & \nodata& \nodata & 0.00&\nodata   &6.07&0.02                     \\
        GY 5  & M7     & -0.96  & \nodata&  1.13  &  0.66  & 10.91   & 5.00 &  -0.06  &$<$5.00&0.07                     \\
       GY 37  & M6     & -1.94  & \nodata&  1.31  &  0.95  & 11.99   & 3.30 &   0.38  &6.64&0.05                     \\
      GY 141  & M8.5   & -2.59  & \nodata&  0.76  &  0.50  & 13.87   & 0.00 &   0.04  &5.93&0.02                     \\
\enddata
\tablecomments{The photometry and spectroscopy used to calculate
extinction values, infrared excesses, effective temperatures and
luminosities comes from the following sources:
Taurus -- 2MASS, Briceno et al. (1999), Briceno et al. (1998), 
Luhman \& Rieke (1998), Kenyon \& Hartmann (1995);
IC 348 -- Luhman (1999), Luhman et al. (1998), Herbig (1998),
Najita et al. (2000);
Sigma Ori -- 2MASS, Bejar et al. (1999) and Bejar et al. (2001);
Upper Sco -- 2MASS, Ardila, Mart{\'\i}n, \& Basri (2000)
Ophiuchus -- Wilking, Greene, \& Meyer (1999), Luhman \& Rieke (1999),
Barsony et al. (1997).}
\end{deluxetable}

\begin{deluxetable}{lccc}
\tablecaption{Velocities \& Lithium
\label{hires_vel}}
\tablewidth{0pt}
\tablehead{
\colhead{object} & \colhead{V$_r$} & \colhead{$v\sin i$} & \colhead{EW(Li $\lambda$6708)}}
\startdata
CIDA 13 &  15.9 $\pm$ 0.5 & \nodata & $<$0.1 \\  
CIDA 14 & 14.0 $\pm$ 0.7 & 11.8 $\pm$ 1.8 & 0.51\\
FN Tau & 14.9 $\pm$ 0.4 & 8.5 $\pm$ 0.8 & 0.52\\
FP Tau & 16.8 $\pm$ 2.0 & 34.2$^a$ $\pm$ 0.8 & 0.58\\
Haro 6-28 &  16.8 $\pm$ 0.9 & \nodata & 0.6 \\ 
LkCa 1 & 9.0 $\pm$ 1.8 & 34.4$^a$ $\pm$ 0.6 & 0.55\\
LkHa 358 (ESI) &  8.9 $\pm$ 4.6 & \nodata & 0.4 \\ 
LkHa 358 (HIRES) & 17.9 $\pm$ 3.1 & 20.4 $\pm$ 1.1 & 0.58\\
MHO-4 &  19.1 $\pm$ 0.8 & \nodata &  0.5 \\
MHO-5 &  12.3 $\pm$ 1.2 & \nodata &  0.5 \\
MHO-6 &  13.6 $\pm$ 0.7 & \nodata &  0.5 \\
MHO-7 &  16.9 $\pm$ 0.8 & \nodata &  0.5 \\
MHO-8 (ESI) &  15.9 $\pm$ 0.9 & \nodata &  0.5 \\
MHO-8 (HIRES) & 15.3 $\pm$ 1.5 & 16.7 $\pm$ 2.4 & 0.57\\
MHO-9 &  13.8 $\pm$ 0.5 & \nodata &  0.6 \\
V410 Anon 13 (ESI) &  20.4 $\pm$ 0.9 & \nodata & 0.5 \\  
V410 Anon 13 (HIRES) & 15.1 $\pm$ 1.3 & 9.8 $\pm$ 3.0 & 0.65:\\
V410 Xray 3 &  14.6 $\pm$ 0.9 & \nodata &   0.5 \\
V410 Xray 5 &  14.7 $\pm$ 0.4 & \nodata &   0.6 \\
V927 Tau & 16.5 $\pm$ 0.6 & 13.3 $\pm$ 1.3 & 0.33\\
IC348-165 &  8.8 $\pm$ 0.6 & \nodata &   0.5 \\
IC348-173 &  13.9 $\pm$ 2.4 & \nodata &   0.5 \\
IC348-205 &  15.7 $\pm$ 3.4 & \nodata &  0.8 \\
IC348-336 &  13.5 $\pm$ 3.0 & \nodata & \nodata \\ 
IC348-363 &  9.5 $\pm$ 1.7 & \nodata & \nodata \\ 
IC348-382 &  9.1 $\pm$ 0.6 & \nodata & \nodata \\ 
IC348-407 &  7.9 $\pm$ 2.6 & \nodata & \nodata \\ 
IC348-415 & 14.7  $\pm$ 2.5 & \nodata & 1.1: \\ 
IC348-454 &  10.5 $\pm$ 1.8 & \nodata & 0.6 \\ 
IC348-478 &  10.8 $\pm$ 1.4 & \nodata & \nodata\\ 
SOri-12 &  29.8 $\pm$ 0.7 & \nodata & 0.6 \\ 
SOri-17 &  19.66 $\pm$ 1.7 & \nodata &  0.8: \\
SOri-25 (ESI) & 30.06 $\pm$ 2.5 & \nodata &  0.6 \\
SOri-25 (HIRES) & 29.6 $\pm$ 2.3 & 9.4 $\pm$ 1.0 & \nodata \\
SOri-29 &  27.1 $\pm$ 1.6 & \nodata &  0.6 \\
SOri-40 &  32.5 $\pm$ 3.3 & \nodata &  0.5 \\
SOri-45 &  22.4 $\pm$ 5.3 & \nodata &  \nodata \\
SOri-46 &  55.8 $\pm$ 2.9 & \nodata &  \nodata \\
UScoCTIO 66 &  -4.4 $\pm$ 0.6 & \nodata & 0.6 \\  
UScoCTIO 75 &  -5.6 $\pm$ 1.1 & \nodata &   0.6 \\
UScoCTIO 85 &  -24.6 $\pm$ 0.7 & \nodata &  $<$0.1 \\
UScoCTIO 100 &  -8.9 $\pm$ 0.6 & \nodata & 0.6 \\ 
UScoCTIO 109 &  -3.8 $\pm$ 0.7 & \nodata & 0.6 \\ 
UScoCTIO 121 &  -38.9 $\pm$ 1.0 & \nodata & $<$0.3 \\ 
UScoCTIO 128 &  -3.0  $\pm$ 1.6 & \nodata & 0.5: \\ 
UScoCTIO 132 &  -8.2 $\pm$ 1.1 & \nodata &  $<$0.4: \\
GY 5 & -6.3 $\pm$ 1.9 & 16.8 $\pm$ 2.7 & 0.5\\
GY 37 & \nodata & \nodata & \nodata \\
GY 141 & \nodata & \nodata & \nodata \\
\enddata
\tablecomments{Velocities in km s$^{-1}$, EW in \AA.  Null values
indicate that measurements could not be made due to low continuum
S/N, or, in the case of $v\sin i$, could not be made due
to the lower resolution of the ESI spectrograph.\\
$^a$Possible spectroscopic binary.}
\end{deluxetable}

\begin{deluxetable}{lcccccccccccccc}
\tabletypesize{\footnotesize}
\rotate
\tablewidth{0pt}
\tablecaption{Emission Lines
\label{emlines}}
\tablehead{
\colhead{object} & \colhead{\ha} & \colhead{\ha} & \colhead{H$\beta$} & \colhead{H$\gamma$} & \colhead{H$\delta$} & \colhead{He I} & \colhead{[OI]} & \colhead{[OI]} & \colhead{Ca II} & \colhead{Ca II} & \colhead{Ca II} & \colhead{accretor?} & \colhead{instrument}\\
\colhead{} & \colhead{10\% width} & \colhead{} & \colhead{} & \colhead{} & \colhead{} & \colhead{$\lambda$5876} & \colhead{$\lambda$6300} & \colhead{$\lambda$6363} & \colhead{$\lambda$8498$^a$} & \colhead{$\lambda$8542$^a$} & \colhead{$\lambda$8662$^a$} & \colhead{} & \colhead{}}
\startdata
CIDA 13$^c$ & 77 & -1.6 & -2.0 & -2.0 & -2.0 & -0.4 & $<$-0.1 & $<$-0.1 & $<$-0.1 & $<$-0.6 & $<$-0.6 & n & ESI\\
CIDA 14 & 289 & -34 & \nodata & \nodata & \nodata & \nodata & \nodata & $<$-0.1 & $<$-0.2 & \nodata & $<$-0.1 & y & HIRES\\
FN Tau & 195 & -22 & \nodata & \nodata & \nodata & \nodata & \nodata & $<$-0.2 & -0.9 & \nodata & -0.7 & n & HIRES\\
FP Tau & 418 & -32 & \nodata & \nodata & \nodata & \nodata & \nodata & $<$-0.1 & -0.1 & \nodata & -0.1 & y & HIRES\\
Haro 6-28 & 347 & -48 & -30 & -25 & -25 & -2.2 & -1.2 & -0.3 & -0.2 & -0.3 & -0.1 & y & ESI\\
LkCa 1 & 178 & -3.9 & \nodata & \nodata & \nodata & \nodata & \nodata & $<$-0.1 & -0.1 & \nodata & $<$-0.1 & n & HIRES\\
Lk\ha 358 & 502 & -63 & -22$^b$ & \nodata & \nodata & -1.9 & -15 & -4.1 & -6.3 & -6.6 & -5.1 & y & ESI\\
Lk\ha 358 & 477 & -62 & \nodata & \nodata & \nodata & \nodata & \nodata & -2.2 & -9.1 & \nodata & -7.8 & y & HIRES\\
MHO-4 & 116 & -43 & -56 & -35 & -25 & -4.3 & 1.2: & $<$-0.1 & -0.4 & -0.5 & -0.1 & n & ESI\\
MHO-5 & 154 & -60 & -44 & -29 & -19 & -3.3 & -4.0 & -0.9 & -0.4 & -0.5 & -0.1 & y & ESI\\
MHO-6 & 309 & -25 & -13 & -11 & -8.6 & -1.7 & -1.4 & -0.3 & $<$-0.1 & $<$-0.1 & $<$-0.1 & y & ESI\\
MHO-7 & 116 & -9.0 &-7.5 & -3.6 & -3.0 & -0.5 & $<$-0.1 & $<$-0.1 & $<$-0.1 & $<$-0.1 & $<$-0.1 & n & ESI\\
MHO-8 & 115 & -18 & -17 & -11 & -5.5 & -1.4 & $<$-0.1 & $<$-0.1 & $<$-0.2 & $<$-0.1 & $<$-0.1 & n & ESI\\
MHO-8 & 128 & -14 & \nodata & \nodata & \nodata & \nodata & \nodata & $<$-0.1 & $<$-0.2 & \nodata & $<$-0.2 & n & HIRES\\
MHO-9 & 116 & -3.5 & -4.3 & -2.8 & -2.4 & -0.5 & $<$-0.2 & $<$-0.1 & $<$-0.1 & $<$-0.3 & $<$-0.2 & n & ESI\\
V410 Anon 13 & 270 & -29 & -20$^b$ & -9$^b$ & \nodata & -12.8$^b$ & -0.8 & $<$-0.5 & $<$-0.1 & $<$-0.1 & $<$-0.1 & y & ESI\\
V410 Anon 13 & 248 & -27 & \nodata & \nodata & \nodata & \nodata & \nodata & $<$-0.4 & $<$-0.3 & \nodata & $<$-0.1 & y & HIRES\\
V410 Xray-3 & 116 & -20 & -23 & -18 & -10 & -2.4 & $<$-0.2 & $<$-0.1 & $<$-0.1 & $<$-0.2 & $<$-0.1 & n & ESI\\
V410 Xray-5a & 154 & -19 & -22 & -15$^b$ & \nodata & -2.1 & $<$-0.5 & $<$-0.2 & -0.2 & -0.3 & $<$-0.1 & n & ESI\\
V927 Tau & 290 & -7.0 & \nodata & \nodata & \nodata & \nodata & \nodata & $<$-0.1 & $<$-0.1 & \nodata & $<$-0.1 & n & HIRES\\
IC348-165 & 347 & -54 & -28 & -24$^b$ & \nodata & -2.4 & -0.3: & -0.2 & -0.3: & -0.4 & -0.4: & y & ESI\\
IC348-173 & 347 & -86 & -50 & -10$^b$ & -35$^b$ & -7.5 & -2.2: & -0.4 & -0.8 & -1.6 & -0.7 & y & ESI\\
IC348-205 & 270 & -105 & -74$^b$ & \nodata & \nodata & $<$-1.5$^b$ & $<$-0.5 & $<$-0.7 & -1.1 & -1.8 & -0.9 & y & ESI\\
IC348-336 & 309 & -121 & -42$^b$ & -20$^b$ & \nodata & -6.4$^b$ & $<$-2.8: & $<$-1.1: & -0.2 & -0.4 & $<$-0.1 & y & ESI\\
IC348-363 & 117 & -13 & \nodata & \nodata & \nodata & \nodata & $<$-21: & $<$-1.0 & $<$-0.1 & $<$-0.1 & $<$-0.1 & n & ESI\\
IC348-382 & 232 & -70 & \nodata & \nodata & \nodata & \nodata & \nodata & \nodata & -0.6 & -0.9 & -0.3: & y & ESI\\
IC348-407 & 155 & -24 & \nodata & \nodata & \nodata & \nodata & \nodata & \nodata & $<$-0.7 & -0.7: & $<$-1.1 & n & ESI\\
IC348-415 & 272 & -152 & -40$^b$ & \nodata & \nodata & \nodata & $<$-4.0:$^b$ & $<$-1.3 & -2.8 & -4.7 & -2.6 & y & ESI\\
IC348-454 & 193 & -23 & -32$^b$ & -18$^b$ & \nodata & \nodata & $<$-3.5: & $<$-3.9 & $<$-0.1 & $<$-0.1 & $<$-0.2 & n & ESI\\
IC348-478 & 117 & -22 & -40$^b$ & \nodata & \nodata & \nodata & $<$-9.6: & $<$-1.5 & $<$-0.1 & $<$-0.2 & $<$-0.1 & n & ESI\\
SOri-12 & 116 & -9 & -9.5 & -5.5$^b$ & \nodata & \nodata & $<$-1.1: & $<$-0.3 & $<$-0.1 & $<$-0.1 & $<$-0.1 & n & ESI\\
SOri-17 & 126 & -4.8 & -5.0$^b$ & \nodata & \nodata & \nodata & $<$-3.5: & $<$-2.2: & $<$-0.1 & $<$-0.1 & $<$-0.1 & n & ESI\\
SOri-25 & 156 & -44 & -50$^b$ & -36$^b$ & \nodata & -6.0 & $<$-1.0 & $<$-0.5 & -0.2: & -0.2 & $<$-0.2 & n & ESI\\
SOri-25 & 94: & -36$^b$ & \nodata & \nodata & \nodata & \nodata & \nodata & \nodata & $<$-0.5 & \nodata & $<$-0.25 & n & HIRES\\
SOri-29 & 154 & -15 & -21 & -16$^b$ & \nodata & $<$-1.5 & $<$-0.2 & $<$-0.1 & $<$-0.1 & $<$-0.1 & $<$-0.2 & n & ESI\\
SOri-40 & 194 & -16 & -12 & \nodata & \nodata & \nodata & $<$-1.0: & $<$-2.3: & $<$-0.1 & $<$-0.3 & $<$-0.2 & n & ESI\\
SOri-45$^c$ & 78: & -21$^b$ & -16$^b$ & \nodata & \nodata & \nodata & \nodata & \nodata & $<$-0.2 & $<$-0.3 & $<$-1.4 & n & ESI\\
SOri-46$^c$ & 194 & -14 & -9$^b$ & \nodata & \nodata & \nodata & \nodata & \nodata & $<$-0.4 & $<$-0.4 & $<$-0.6 & n & ESI\\
UScoCTIO-66 & 116 & -6.0 & -8.8 & -8.0 & -5.5 & -1.0 & $<$-0.3 & $<$-0.1 & $<$-0.1 & $<$-0.1 & $<$-0.1 & n & ESI\\
UScoCTIO-75 & 231 & -15 & -19 & -16 & -10 & -1.3 & $<$-0.1 & $<$-0.1 & $<$-0.1 & $<$-0.1 & $<$-0.1 & n & ESI\\
UScoCTIO-85$^c$ & 154 & -7.0 & -14 & -18$^b$ & -5.0$^b$ & -0.9 & $<$-0.2 & $<$-0.2 & $<$-0.1 & $<$-0.2 & $<$-0.2 & n & ESI\\
UScoCTIO-100 & 154 & -12 & -23 & -20 & -12 & -2.0 & $<$-0.1 & $<$-0.1 & $<$-0.1 & $<$-0.1 & $<$-0.1 & n & ESI\\
UScoCTIO-109 & 116 & -10 & -22 & -12 & -11$^b$ & -2.5 & $<$-0.3 & $<$-0.1 & $<$-0.2 & $<$-0.1 & $<$-0.1 & n & ESI\\
UScoCTIO-121$^c$ & 116 & -7.3 & -25$^b$ & \nodata & \nodata & \nodata & $<$-0.4: & $<$-1.2 & $<$-0.1 & $<$-0.1 & $<$-0.2 & n & ESI\\
UScoCTIO-128 & 193 & -60 & -50 & -50$^b$ & -30$^b$ & -10$^b$ & $<$-0.3 & $<$-0.6 & -0.3 & -0.6 & $<$-0.2 & n & ESI\\
UScoCTIO-132$^c$ & 116 & -4 & \nodata & \nodata & \nodata & \nodata & $<$-3.0: & $<$-0.6 & $<$-0.2 & $<$-0.3 & $<$-0.4 & n & ESI\\
GY 5 & 177 & -17 & \nodata & \nodata & \nodata & \nodata & \nodata & -0.4: & $<$-0.2 & \nodata & $<$-0.1 & n & HIRES\\
GY 37 & 197: & -20$^b$ & \nodata & \nodata & \nodata & \nodata & \nodata & \nodata & $<$-0.4 & \nodata & $<$-0.4 & n & HIRES\\
GY 141 & 88: & -32$^b$ & \nodata & \nodata & \nodata & \nodata & \nodata & \nodata & $<$-3.0 & \nodata & $<$-17 & n & HIRES\\
\tableline
Gl 406 & 113 & -7.1 & -13 & -15 & -16 & -0.9$^d$ & $<$-0.1 & $<$-0.1 & $<$-0.1 & -0.1 & $<$-0.1 & n & ESI\\
LHS 2351 & 136 & -3.9 & -11 & -9.2 & -9.1$^b$ & -0.6$^d$ & $<$-0.1 & $<$-0.1 & $<$-0.1 & -0.2: & $<$-0.2 & n & ESI\\
\enddata
\tablecomments{Line measurements are equivalent widths in {\AA}, unless otherwise noted.  H$\alpha$ 10\% widths are in km s$^{-1}$.  Colons indicate uncertain measurements due to poor S/N, or imperfect sky subtraction at the [OI] lines.  Accretors have been selected based on \ha characteristics, as described in section 4.  Gl406 and LHS 2351, main sequence dwarfs with spectral types M5.5 and M7Ve, respectively, are included for comparison.\\
$^a$ Ca II EW of emission component; measured against absorption
pseudo-continuum when photospheric absorption component is not completely
filled in by the emission.\\
$^b$ Continuum not detected at line position.\\
$^c$ Most likely not a member of the association (see text).\\
$^d$ Measured from a pseudo-continuum within absorption wing of Na I line.}
\end{deluxetable}

\begin{deluxetable}{lcccccc}
\tablecaption{ESI Accretor Sample
\label{veiling1}}
\tablewidth{0pt}
\tablehead{
\colhead{object} & \colhead{veiling template} & \colhead{r$_{5500}$} & \colhead{r$_{6200}$} & \colhead{r$_{7100}$} & \colhead{r$_{8900}$} & \colhead{log $\mdot \; (\msunyr)$}}
\startdata
Haro 6-28 & TWA 15A, CIDA 13$^a$ & 0.3 & 0.2 & 0.1 & 0.0 & -8.7$^b$\\
Lk\ha 358 & LkCa 7 & \nodata & 1.0: & 1.0: & 1.0: & -8.5$^b$\\
MHO-5 & V410 Xray-3 & 0.0 & 0.1 & 0.0 & 0.0 & -10.8\\
MHO-6 & MHO-7 & 0.0 & 0.1 & 0.1 & 0.1 & -10.3\\
V410 Anon13 & V410 Xray-3 & 0.0 & 0.1 & 0.0 & 0.1 & -11.3\\
IC348-165 & MHO-8 & 0.0 & 0.0 & 0.0 & 0.0 & -10\\
IC348-173 & MHO-7 & 0.0 & 0.1 & 0.0 & 0.0 & -10\\
IC348-205 & MHO-8 & 0.0: & 0.1 & 0.0 & 0.1 & -10\\
IC348-336 & MHO-7 & \nodata & 0.2: & 0.1 & 0.1 & -10\\
IC348-382 & V410 Xray-3 & \nodata & \nodata & 0.2: & 0.2: & -10.8\\
IC348-415 & V410 Xray-3 & \nodata & 0.5: & 0.1 & 0.1 & -9.3\\
\enddata
\tablecomments{Typical veiling errors are $\sim \pm 0.1$.
Colons indicate larger uncertainties due to poor S/N.
Null values are cases where the S/N was so low that
no measurements could be made.  Accretion rates are in
$\msunyr$, all determined from \ha modeling except for $^b$,
which were determined from the veiling.\\
$^a$ Veiling template was the average of these two stars,
whose spectral types are M1.5 and M3.5, respectively.}
\end{deluxetable}

\begin{deluxetable}{lccccc}
\tablecaption{Veiling - HIRES Accretor Sample
\label{veiling2}}
\tablewidth{0pt}
\tablehead{
\colhead{object} & \colhead{template} & \colhead{r$_{6450}$} & \colhead{r$_{7100}$} & \colhead{r$_{8700}$} & \colhead{log $\mdot \; (\msunyr)$}}
\startdata
CIDA 14 & MHO-8 & -0.1 & 0.1 & -0.1 & -10.3\\
FP Tau & LkCa 1 & 0.0 & 0.2 & -0.1 & $<$-9$^a$\\
Lk\ha 358 & LkCa 7 & 0.4 & 0.4 & 0.4 & -8.5$^a$\\
V410 Anon 13 & MHO-8 & \nodata & 0.1 & 0.0 & -11.3\\
\enddata
\tablecomments{Typical veiling errors are $\sim \pm 0.1$.
Null value indicates no S/N in the object spectrum around
that wavelength.  Accretion rates are in $\msunyr$,
all determined from \ha modeling except for $^a$, which were
determined from the veiling.}
\end{deluxetable}

\begin{deluxetable}{lcccccc}
\tablecaption{Parameters for \ha Model Comparisons
\label{model_param}}
\tablewidth{0pt}
\tablehead{
\colhead{object} & \colhead{$M_* \; (M_{\odot})$} & \colhead{$R_* \; (R_{\odot})$} & \colhead{$R_{mag} \; (R_*)$} & \colhead{$T_{max} \; (K)$} & \colhead{$i \; (^\circ)$} & \colhead{log $\mdot \; (\msunyr)$}}
\startdata
MHO-5 & 0.05 & 0.5 & 2.2-3 & 12,000 & 45 & -10.8\\
MHO-6 & 0.15 & 1.0 & 2.2-3 & 12,000 & 60 & -10.3\\
CIDA 14 & 0.15 & 1.0 & 2.2-3 & 12,000 & 50 & -10.3\\
IC348-165 & 0.15 & 1.0 & 2.2-3 & 12,000 & 89 & -10.0\\
IC348-173 & 0.15 & 1.0 & 2.2-3 & 12,000 & 80 & -10.0\\
IC348-205 & 0.15 & 1.0 & 2.2-3 & 12,000 & 80 & -10.0\\
IC348-336 & 0.15 & 1.0 & 2.2-3 & 12,000 & 75 & -10.0\\
IC348-415 & 0.05 & 0.5 & 2.2-3 & 12,000 & 45 & -9.3\\
IC348-382 & 0.05 & 0.5 & 2.2-3 & 12,000 & 55 & -10.8\\
\enddata
\tablecomments{$R_{mag}$ is the inner and outer radius of
the magnetospheric accretion flow at the disk surface.  $T_{max}$ is
the maximum value of the temperature distribution of the flow.
The inclination angle is measured from the axis of rotation
of the system.}
\end{deluxetable}

\begin{figure}
\plotone{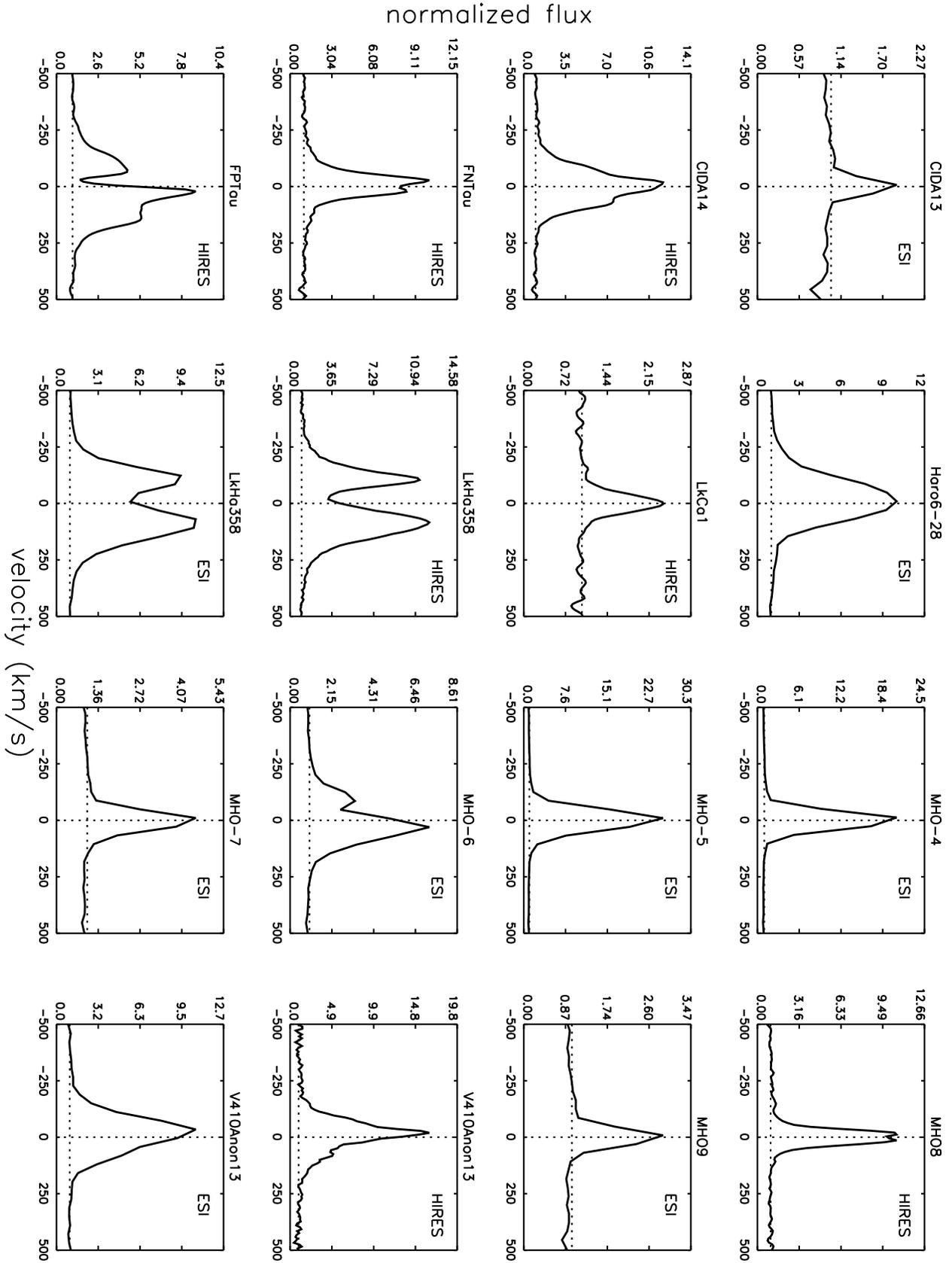}
\caption{Continuum-normalized \ha profiles; see Table~\ref{emlines}
for further information.
\label{halpha_obs}}
\end{figure}

\begin{figure}
\plotone{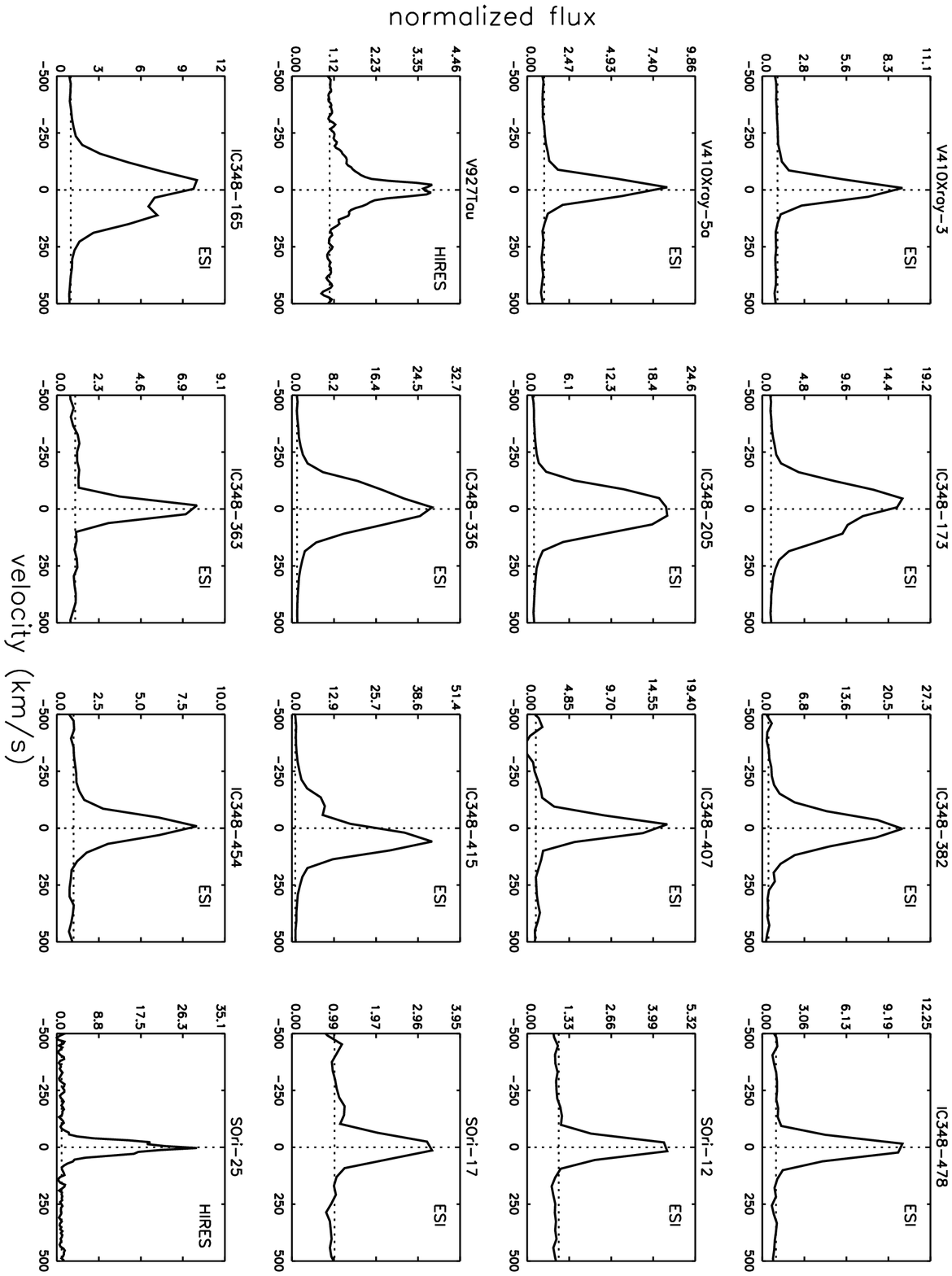}
\addtocounter{figure}{-1}
\caption{Continued.}
\end{figure}

\begin{figure}
\plotone{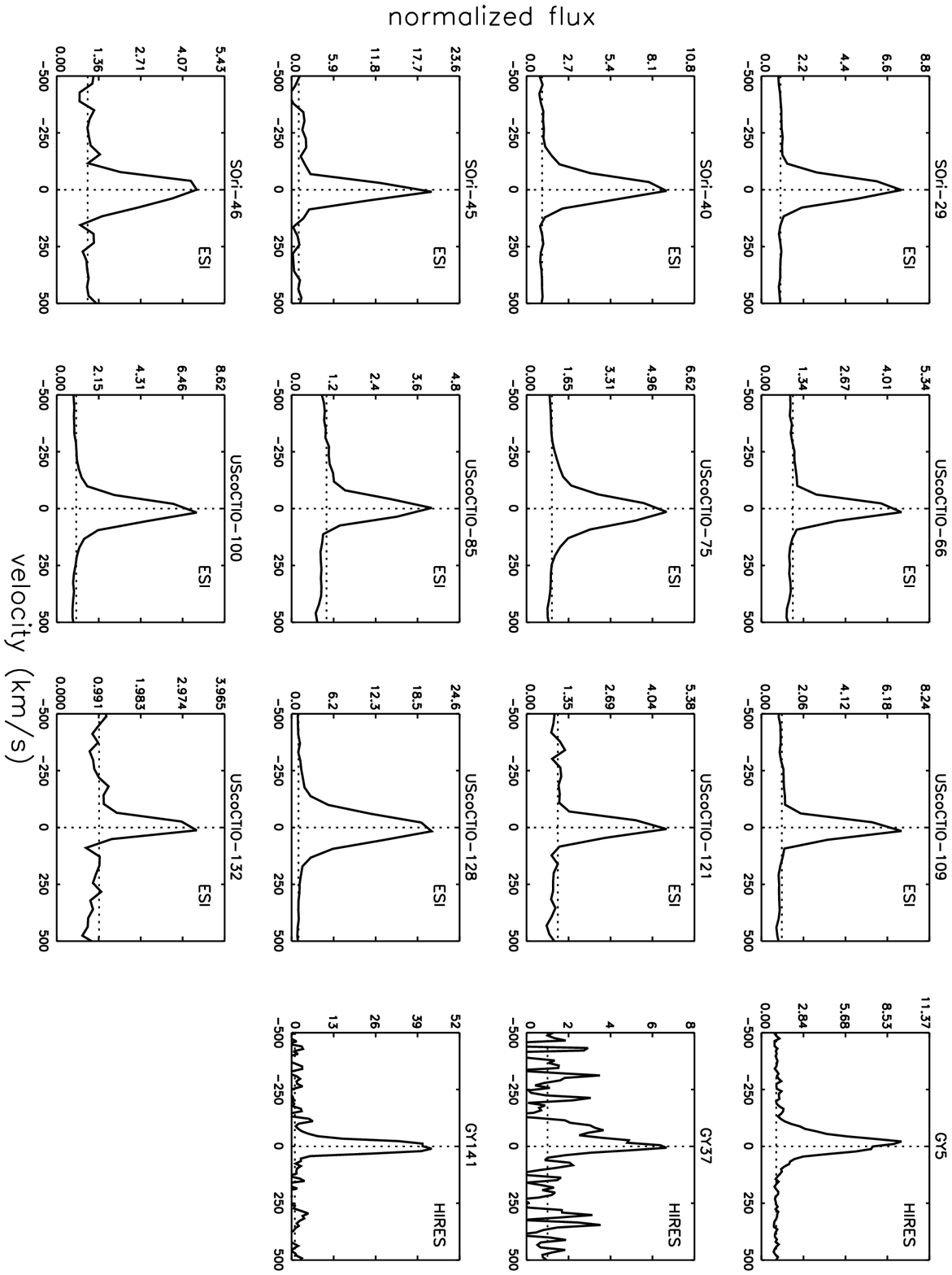}
\addtocounter{figure}{-1}
\caption{Continued.}
\end{figure}

\end{document}